\RequirePackage{lineno} 

\documentclass[preprint2]{aastex}

\usepackage{wrapfig}

\shorttitle{Solar radio burst statistics}
\shortauthors{Saint-Hilaire et al.}

\begin{document}

\title{A decade of solar Type III radio bursts observed by the Nancay Radioheliograph 1998--2008}

\author{P. Saint-Hilaire\altaffilmark{1}, N. Vilmer\altaffilmark{2}, and A. Kerdraon\altaffilmark{2}}
\affil{Space Sciences Laboratory, University of California, Berkeley, CA 94720, USA}
\affil{LESIA, Observatoire de Paris, CNRS, UPMC, Universit\'e  Paris-Diderot 5 place Jules Janssen, 92195 Meudon, France}

\email{shilaire@ssl.berkeley.edu}

\begin{abstract}
	We present a statistical survey of almost 10 000 radio Type III bursts observed by the Nan\c{c}ay Radioheliograph from 1998 to 2008, covering nearly a full solar cycle.
	In particular, sources sizes, positions, and fluxes were examined.
	We find an east-west asymmetry in source positions which could be attributed to a 6$\pm$1$^{\circ}$ eastward tilt of the magnetic field, 
	that source FWHM sizes $s$ roughly follow a solar-cycle averaged distribution $\frac{dN}{ds} \approx 14 \, \nu^{-3.3} s^{-4}$ arcmin$^{-1}$ day$^{-1}$,
	and that source fluxes closely follow a solar-cycle averaged $\frac{dN}{dS_{\nu}} \approx 0.34 \, \nu^{-2.9} S_{\nu}^{-1.7}$ sfu$^{-1}$ day$^{-1}$ distribution (when $\nu$ is in GHz, $s$ in arcmin, and $S_{\nu}$ in sfu).
	Fitting a barometric density profile yields a temperature of 0.6 MK, while a solar wind-like ($\propto h^{-2}$) density profile yields a density of 1.2$\times$10$^{6}$ cm$^{-3}$ at an altitude of 1 $R_S$, assuming harmonic emission.
	Finally, we found that the solar-cycle averaged radiated Type III energy could be similar in magnitude to that radiated by nanoflares via non-thermal bremsstrahlung processes,
	and we hint at the possibility that escaping electron beams might carry as much energy away from the corona as is introduced into it by accelerated nanoflare electrons.
\end{abstract}


%++++++++++++++++++++++++++++++++++++++++++++++++++++++++++++++++++++++++++++++++++++++++++++++++++++++++++++++++++++++++++++++++++++++++++++++++++++++++++++++++++++++++++++++++++++++++++
%++++++++++++++++++++++++++++++++++++++++++++++++++++++++++++++++++++++++++++++++++++++++++++++++++++++++++++++++++++++++++++++++++++++++++++++++++++++++++++++++++++++++++++++++++++++++++
%++++++++++++++++++++++++++++++++++++++++++++++++++++++++++++++++++++++++++++++++++++++++++++++++++++++++++++++++++++++++++++++++++++++++++++++++++++++++++++++++++++++++++++++++++++++++++
\section{Introduction}
	
	Particle acceleration events in the quasi-collisionless plasma of the solar corona create supra-thermal electron beams that propagate along magnetic field lines,
	either away from the Sun into interplanetary space (on so-called ``open'' field lines), or downwards into the chromosphere (towards the footpoints of magnetic loops).
	Higher-energy electrons race ahead of the lower energy ones, creating a bump-in-tail instability in the particle distribution.
	Landau resonance with the unstable electron beams creates Langmuir waves, which are believed to undergo non-linear wave-wave interaction and produce electromagnetic emissions at the local plasma frequency or its harmonic.
	As the electron beam propagates to higher or lower altitudes \citep[at speeds of $\approx c/3$ in the low corona, to $\approx c/10$ at 1 AU,][]{Poquerusse1996}, 
	and hence to lower or higher densities and plasma frequencies, there is a drift in the frequency of emitted radiation. 
	This coherent plasma emission is generally called a Type III radio burst.
	Type IIIs have been observed from frequencies as high as $\approx$1 GHz at the bottom of the corona to 30 kHz at $\approx$1 AU, and even lower further out.
	Taking only bursts drifting from high to low frequencies, \citet{Alvarez1973} have fitted bursts in the 74 to 550 MHz range by the relation 
	$\frac{d\nu}{dt}\approx$ -0.01 $\nu^{1.84}$  MHz s$^{-1}$.

	\citet{Nita2002} have investigated the peak flux distributions of 40 years of spatially-integrated solar radio burst data, from 0.1 GHz to 37 GHz.
	They have found that the peak flux density distribution of events, $dN/dS_{\nu}$ followed a power-law with a negative index $\approx$1.8, similar to that found in many X-ray studies \citep[for recent work on the topic, see e.g.][]{Hannah2008}.
	%\citep[for a recent review, see][]{Aschwanden2010}.
	
	In this work, we have gathered the solar Type III peak flux densities, source sizes and positions at different frequencies, observed by the Nan\c{c}ay Radioheliograph (NRH) over the period 1998-Jan-01 to 2008-Apr-01, 
	i.e. covering nearly a full solar cycle.
	
	In Section \ref{sect:selection}, we describe the data selection process and some of its limitations.
	We describe the observations in Section \ref{sect:observations} 
	and discuss them in Section \ref{sect:discussion}.
	Finally, in Section \ref{sect:ccl}, we summarize our results and present some conclusions to this study.

%++++++++++++++++++++++++++++++++++++++++++++++++++++++++++++++++++++++++++++++++++++++++++++++++++++++++++++++++++++++++++++++++++++++++++++++++++++++++++++++++++++++++++++++++++++++++++
%++++++++++++++++++++++++++++++++++++++++++++++++++++++++++++++++++++++++++++++++++++++++++++++++++++++++++++++++++++++++++++++++++++++++++++++++++++++++++++++++++++++++++++++++++++++++++
%++++++++++++++++++++++++++++++++++++++++++++++++++++++++++++++++++++++++++++++++++++++++++++++++++++++++++++++++++++++++++++++++++++++++++++++++++++++++++++++++++++++++++++++++++++++++++
\section{Data selection and caveats} \label{sect:selection}

	We selected our events from the list of solar radio bursts published by the {\it National Oceanographic and Atmospheric Administration's National Geophysical Data Center}\footnote{ftp://ftp.ngdc.noaa.gov/STP/SOLAR\_DATA/SOLAR\_RADIO/SPECTRAL/} (NOAA/NGDC), 
	from 1998 January 1 to 2008 April 1, covering almost a full solar cycle.
	We removed all bursts that were outside the observing times and spectral range ($\sim$150--450 MHz) of the {\it Nan\c{c}ay Radioheliograph} \citep[NRH,][]{Kerdraon1997}.
	We have not gone further than April 2008 because NRH started observing the Sun with a different set of frequencies, and we wanted a consistent set of frequencies for our statistical study.
	The NOAA list contains information on individual radio bursts, such as reporting station, start and end times and frequencies, as well as spectral type and intensity.
	This list is compiled from reports, often generated manually, by observers from various ground stations around Earth (Culgoora, Ismiran, Learmonth, Ondrejov, Palehua, Sagamore Hill, San Vito, Bleien, etc.),
	using spectrometers covering varying bands, from as low as $\sim$30 MHz to as high as a few GHz.
	Hence, certain characteristics such as start and end frequencies, as well as burst intensity, are somewhat subject to both individual spectrometer sensitivity and individual observer's perception.
	In cases where the same event was reported more than once (by different observatories), only the first such report was kept, and any extraneous ignored.
	We kept only Type III radio burst reports within the times ($\approx$8:30 to 15:30 UT, varying during the year) and the spectral band (150--432 MHz) where the NRH was observing the Sun.
	Groups of decimetric Type IIIs do not last much longer than a few minutes (individual ones last about one second).
	Different groups of Type IIIs, even hours apart, are however sometimes bundled together in a single report.
	Moreover, certain reported time interval are clearly erroneous, probably due to data entry error.
	We have therefore ignored all Type III reports longer than 10 minutes, (which constituted $\sim$14\% of Type III reports), and were left with 8931 Type III burst reports from 1998 to 2008.
	%We removed events which lasted more than 10 minutes.
	%This cutoff was chosen for easier data handling, and because this study was mainly concerned with Type IIIs. 
	%We have collected 10483 solar radio bursts (Type IIIs: 8931, Type IIs: 173, Type Is: 372, DCIMs: 668, Type Vs: 153, and unclassified: 186).
	%For this study, we have decided to concentrate on Type III bursts exclusively, and have discarded the rest.

		\begin{figure*}[ht!]
		\centering
		\includegraphics[width=16.6cm]{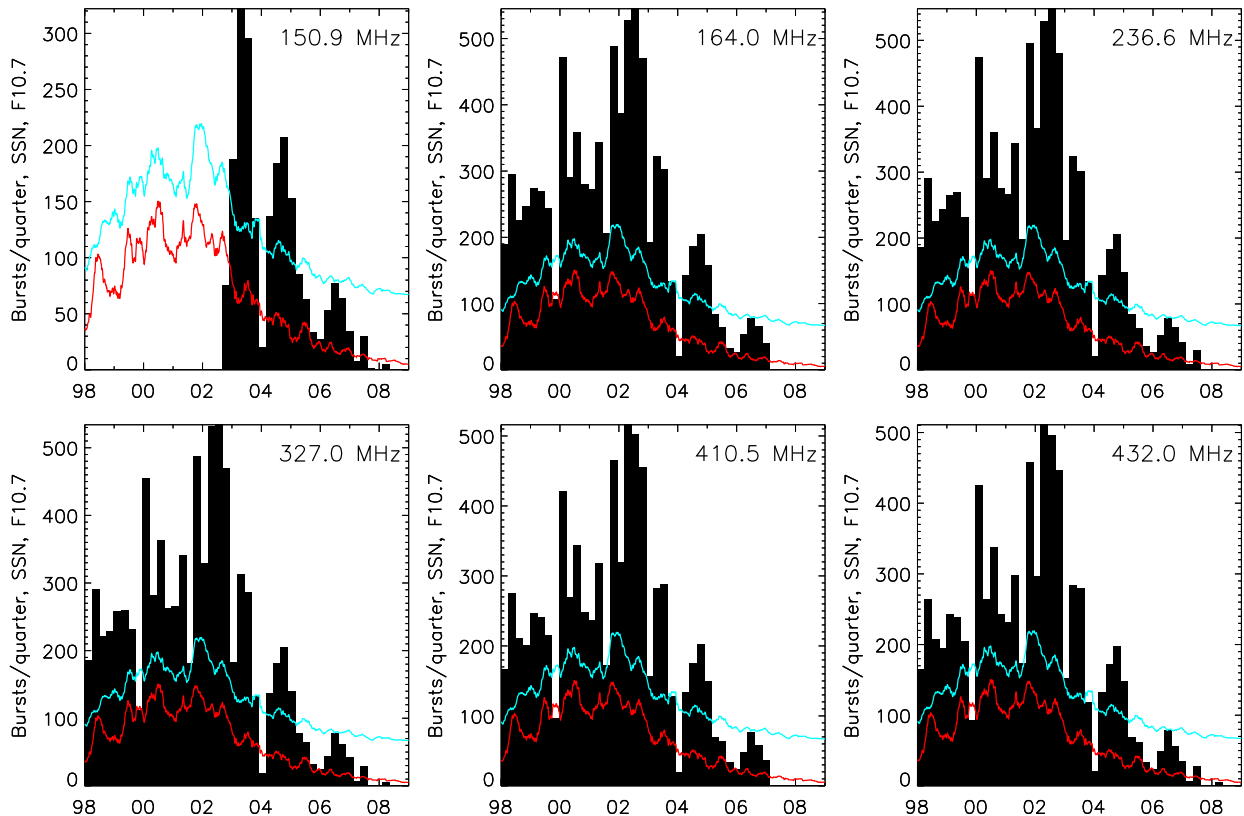}
		\caption{
			{\it Black}: Time series of the quarterly number of NOAA-reported solar radio bursts, at times and frequencies when and where NRH was observing.
			{\it Red}: Sun Spot Number.
			{\it Blue}: F10.7 index (in sfu).
			Three-month time bins or smoothing window were used for all displayed data.
			NRH was not observing at 150.9 MHz before mid-2002, and was not operating from November 2003 to January 2004 (see text in \S\ref{sect:selection} for more details).
		}
		\label{fig:histo_frq_time_distribution}
		\end{figure*}

	The NRH actually produces images at up to 1/8 s temporal resolution, but these are stored on magnetic tapes, and each individual day of data must be retrieved manually.
	The 10-second data is stored in a file system, and it is therefore much easier and faster to process.
	This is the reason why it is used in this study.
	This coarse temporal resolution leads to weak and/or short-duration (such as second-long Type IIIs) bursts being drowned in the solar background.
%	Hence, we have tried to fit 2D gaussians to every radio burst, and kept those sources which yielded ``reasonable'' fitted source sizes, 
%	which we defined as having gaussian FWHM less than half a solar radius (see Section~\ref{sect:size_distribution}),
%	and which had a peak brightness temperature above 10 MK, a value which should 
	Hence, we have decided to retain only well-determined Type III bursts for this study: those which had peak brightness temperature of 10 MK or above, well above any quiet Sun value.
	%This is the case for 90.0, 93.8, 95.4, 9g, 86.3, and 84.1\% of bursts that we had at 150.9, 164, 236.6, 327, 410.5, and 432 MHz.
	This is the case for 91.4\% of cases at 164 MHz, down to 26.2\% of cases at 432 MHz.
	The difference is easily explained by the fact that Type III bursts tend to have lower brightness temperatures at 432 than at 164 MHz (see also \S\ref{sect:fluxes}).
	A priori, the NGDC-reported burst spectral band is not necessarily a good indicator of the actual bandwidth of the Type III burst.
	(NRH, possessing $\approx$44 dishes, is much more sensitive than the single-dish spectrometers that typically report to NGDC).
	On the other hand, we have empirically observed that taking only bursts either at frequencies for which the peak brightness temperature in the map is above $\approx$10 MK, 
	or within the NOAA-reported spectral range, lead to very similar datasets.

	The 10-second data used in this study also leads to cases where several individual bursts are being bundled together into a single event.
	Hence, in this work, a ``Type III event'' is often ``a group of Type IIIs within the same 10-second interval''.

	Frequencies 150.9, 164, 236.6, 327, 410.5, and 432 MHz (with a 3-dB bandwidth of 0.7 MHz) were used by NRH during our period of interest.
	%As we are mostly concerned with statistical studies in this work, only these six frequencies will be considered.
	Figure~\ref{fig:histo_frq_time_distribution} clearly shows a decrease in radio activity as the solar cycle approaches minimum,
	and the dip in events from 2003 November 5 to 2004 January 25 is due to NRH being off-line while anti-alias antennas were being added to the array.
	During our decade of interest, the 150.9 MHz frequency has gradually replaced the nearby 164 MHz, no longer reserved under French law.
	Hence the observations at 150.9 MHz have started later, with an overlap of a few years with the (now no longer used) 164 MHz.
{\bf	It can be noted that the occurrence rates match qualitatively that of \citet{Lobzin2011}, including an almost $\sim$two-year periodicity, unexplained so far.
}
	Using NRH 10-second images, and a fixed 30'' pixel resolution at available frequencies (this standard coarse resolution, still well below the instrumental resolution at 432 MHz, was chosen for ease of data handling), 
	the following were determined (among other things):
	time of peak flux density, location and brightness temperature of brightest pixel in the source, and 2D gaussian fitting parameters to sources present in the image.
	From the latter, source fluxes can be deduced by computing the gaussian volume.
	It is important to note that each frequency was treated mostly independently from each other, see \S\ref{sect:timeintv} and \ref{sect:alias} for a discussion.

	Paragraphs 2.1 to 2.6 detail some caveats in our selection and data analysis process.

\subsection{Time interval of reports:} \label{sect:timeintv}
	Firstly, as stated earlier, there can be several Type IIIs temporarily close to one another, within the 10-second accumulation of the data at hand, 
	leading to confusion and averaging of their characteristics.
	Secondly, our methods looks only for the strongest burst within the NOAA report's time interval (there can be several inside the time range reported), 
	and does so for each frequency independently.
	This has for consequence that we are missing any other bursts within the reported time interval, 
	and that the burst characteristics derived at different frequencies might not relate to the exact same Type III burst.
	About $\sim$50\% of radio bursts from the same report have hence a time difference greater than 10 s between 164 and 432 MHz).
	This has an obvious effect on the absolute occurrence rates of Type III bursts (they are probably higher than reported here), but we estimate the influence on the power-law shape of their distributions (see \S\ref{sect:observations}) to be minimal.
	%{\bf [I think we could have a more definitive answer with a few Monte-Carlo simulations... Can also try to change the 10 minutes cutoff, and see if it has any effects...]}	
		
\subsection{Alias ambiguity:} \label{sect:alias}

	For bursts occurring before the installation of anti-alias antennas at NRH (between 2003 November 5 and 2004 January 25), there can be an ambiguity in source positions at frequencies above 164 MHz.
	This is because an alias is sometime present in the reconstructed NRH image, and sometimes has the brightest pixel in the map
	(particularly when the real source happens to be beyond the solar limb, but the alias lies on top of the solar disk).
	The distance between the real source and an alias decreases with increasing observing frequency, and any alias is beyond the imaged field of view (and hence no alias ambiguity exists) at frequencies $\lesssim$164 MHz.
	It is usually easy to distinguish between a solar source lasting at least a few minutes and its alias: the alias moves quickly (with time) in a straight line across the solar map.
	But this method is of little use in our study, as we examine events that seldom last longer than a single 10-second frame.
	We will instead use the fact that the true source is likely to be spatially near the sources at other frequencies, assuming the radio bursts spans the necessary frequency range.
	For maps taken at progressively higher frequencies than 164 MHz, we determine whether there is possiblity of an alias in the map in the following manner:
	after locating the source with the brightest pixel, we determine whether there is another source in the map at an appropriate alias position, and with similar peak intensity 
	(within 20\%, which corresponds to a 2 MK difference for the minimal 10 MK burst).
	%{\bf !!!! In fact, I should have subtracted the quiet Sun component (if any) from each source. I can still try to do it.)}
	The ``correct'' source is deemed to be the one whose position is closest to the position of the ``correct'' source in the map of the next lower frequency.
	This simple method is not perfect: in some cases, emission at lower frequencies is simply not there, because the conditions for plasma emission and/or propagation to Earth are not adequate, 
	and hence, at higher frequencies, the method may home in on other sources.

	Before the installation of anti-alias antenna, there was a {\it potential} alias confusion (two sources of similar intensities, at correct alias positions, as described above) in our data 58\% of the time at 432 MHz, progressively lower at lower frequencies, down to 0\% at 164 MHz and below.

	In summary, source locations before January 2004, low frequencies (150.9 and 164 MHz) should be always accurate, while higher frequencies may have errors in some of the positions,
	but we expect the impact on the statistical distributions of burst locations (\S\ref{sect:locations}) to be minimal (see coronal density fittings in \S\ref{sect:discussion} for a confirmation).
	The study of source sizes (\S\ref{sect:size_distribution}) and fluxes (\S\ref{sect:fluxes}) should be unaffected by this issue, as is all the data taken after January 2004.

\subsection{Bursts fluxes:}
	The so-called ``Quiet Sun'' flux is of the order of 10 sfu at NRH frequencies (and brightness temperatures at Sun center of $\sim$0.7 MK at 164 MHz).
	Hence, a lone, 1-sfu, 1-second long Type III burst with area about 1/100 the solar disc (i.e. smeared out to about the size of the NRH beam at 164 MHz, {\bf about 3.2' FWHM}) would produce an additional surface brightness equal to that of the quiet Sun in a 10-second image.
	Successfully fitting a gaussian to such a weak burst becomes difficult, and explains the turnover at low fluxes in the distributions presented in \S\ref{sect:fluxes}.

\subsection{Astrophysical sources:}
	The Crab Nebula, or one of its alias, is the most obvious astrophysical source which can enter the NRH field of view (around June each year).
	Its surface brighteness is similar to the Sun at NRH frequencies, i.e. much lower than the 10 MK threshold used for our selection.
	Hence, we expect such sources to have little impact in our study.

\subsection{Refractive effects:}
	Earth ionospheric refraction can significantly perturb the apparent position of a source in ground-based observatories, sometimes exceeding several arcmins in the 100--300 MHz range \citep[e.g.][]{Bougeret1981,Mercier1986}.
	These effects are mostly random position shifts due to ionospheric gravity waves.
	They are not correlated with the positions of sources on the Sun, and should not affect our statistics.
	We have therefore ignored those effects in the present study.

%\subsection{Dearth of afternoon events:}
%	As is plainly visible in Figure~\ref{fig:histo_time}, there is a dearth of bursts in the afternoon (NRH time zone).
%	This dearth also appears (in similar proportions) in the NOAA list, and is therefore most probably not due to problematic data at NRH.
%	We do not have a clear explanation for this. 
%	As far as we know, there appears to be enough overlap from observatories around these times. 
%	It it possible that observers are less diligent reporting radio bursts during their afternoons?
%	In any case, this peculiarity has little effect on most of the results we are presenting in this work.

%+++++++++++++++++++++++++++++++++++++++++++++++++++++++++++++++++++++++++++++++++++++++++++++++++++++++++++++++++++++
%+++++++++++++++++++++++++++++++++++++++++++++++++++++++++++++++++++++++++++++++++++++++++++++++++++++++++++++++++++++
%+++++++++++++++++++++++++++++++++++++++++++++++++++++++++++++++++++++++++++++++++++++++++++++++++++++++++++++++++++++
\section{Data Analysis} \label{sect:observations}

	%We have a very rich dataset.
	%We will present only what we deemed to be the most interesting plots and correlations,
	%and have not tried to plot and discuss all possible combinations of available variables: source frequency $\nu$, brightness temperature $T_b$, source solid angle $\Omega$, flux $S_{\nu}$, position ($x$,$y$) (or solar longitude and latitude), and their time in the solar cycle or since local midnight.
	%%{\bf (perhaps we should eventually... Another paper?)}

	%This section is divided into four sub-sections.
	%Sub-section~\ref{sect:size_distribution} will deal with the distributions of source sizes, 
	%sub-section~\ref{sect:locations} will deal with source positions,
	%sub-section~\ref{sect:fluxes} will deal with the distribution of source fluxes.
	%And, finally, sub-section~\ref{sect:rocorr} will present the rank-order cross-correlation coefficients obtained for all our burst parameters, for three frequencies.

	We have studied in detail the distribution in spatial sizes (\S\ref{sect:size_distribution}), in positions (\S\ref{sect:locations}), in fluxes (\S\ref{sect:fluxes}),
	and the rank-order cross-correlations between several of the observables (\S\ref{sect:rocorr}).

\subsection{Burst spatial size distribution:} \label{sect:size_distribution}

\begin{figure}[ht!]
		\centering
		\includegraphics[width=7.5cm]{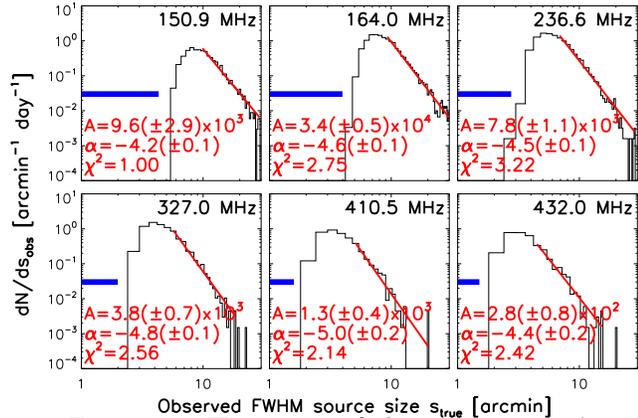}
		\caption{
			Histogram of the radio sources' rms FWHM $s=s_{rms,obs}$, at different frequencies.
			%Contrary to the rest of this work, we have also included events with $\sigma_{rms,obs} \geq R_S/2$.
			%The red curves are fittings $f(r) = \left( \frac{r}{a}\right)^\alpha $ to the the histograms, from the peak of the histogram to higher mean source radius $\sigma$.
			The red curves are fittings $\frac{dN}{ds}= A \, s^\alpha$ to the the histograms, from the bin to the immediate right of the peak in the histogram and above.
	                The blue bar delimits the {\it minimum} NRH point spread function (which can vary by almost a factor 2, depending on the season and time of day).	
		}
		\label{fig:histo_raw_src_sizes_minTb}
		\end{figure}

	Radio sources were fitted with 2D elliptical gaussians. 
	These yield {\it observed} gaussian sizes $\sigma_{a,obs}$ and $\sigma_{b,obs}$ (semi-major and semi-minor $e$-folding lengths), and a tilt angle $\theta_{obs}$ of the semi-major axis with respect to the map x-axis (solar east-west).
	The FWHM values, $s_{a,obs}$ and $s_{b,obs}$, are obtained by multiplication of $\sigma_{a,obs}$ and $\sigma_{b,obs}$  with $2\sqrt{2\ln2} \approx 2.355$.
	In Figure~\ref{fig:histo_raw_src_sizes_minTb}, we plot a histogram of the rms averages $s_{rms,obs}=\sqrt{s_{a,obs}^2 + s_{b,obs}^2}$.
	The roll-over at low source sizes is due to the interferometer beamwidth, which is about 3.2--5.5' FWHM at 164 MHz, proportionaly smaller at higher frequencies,
	and varying with the season and time of day.

\begin{figure}[ht!]
		\centering
		\includegraphics[width=7.5cm]{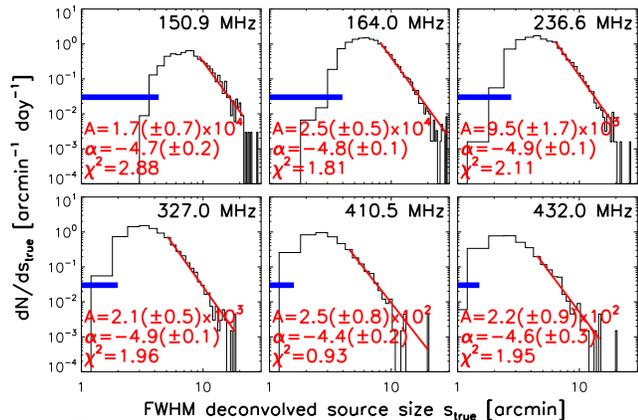}
		\caption{
			As in Figure~\ref{fig:histo_raw_src_sizes_minTb}, but displaying {\it true} (deconvolved) FWHM rms source sizes.
			The red line and labels are power-law fittings to the data.
			The blue bar delimits the {\it minimum} NRH point spread function (which can vary by almost a factor 2, depending on the season and time of day).
		}
		\label{fig:histo_true_src_sizes}
		\end{figure}

	Assuming that the observed source, the interferometer beam, and the true (deconvolved) source were all elliptical gaussians (with different tilt angles),
	we have recovered the true source size (see Appendix~\ref{appendix:deconvolution} for mathematical details), and plotted the results in Figure~\ref{fig:histo_true_src_sizes} and their averages in Table~\ref{tab:srcsize}.
	%Notice that both power-law and gaussian fitting to the distributions are about equally good (or bad).
	Note that some of the largest events (in size) are probably multiple sources, but it was not the purpose of this statistical work to discriminate between such.

			\begin{table}[h!]
			\caption{Mean and standard deviation of observed and deconvolved (or ``true'') source sizes:}
			\centering
			\begin{tabular}{lcc}
				\hline
				Frequency:		&  Observed	&	Deconvolved	\\
				$[$MHz$]$		&  size	[']	&	size [']	\\
				\hline \hline
				150.9			& 10.9$\pm$4.1	& 5.3$\pm$1.8		\\
				164			& 9.7$\pm$3.6	& 4.5$\pm$1.6		\\
				236.6			& 7.0$\pm$2.7	& 3.4$\pm$1.3		\\
				327			& 5.2$\pm$1.9	& 2.5$\pm$1.0		\\
				410.5			& 4.2$\pm$1.6	& 2.1$\pm$0.9		\\
				432			& 3.9$\pm$1.6	& 1.9$\pm$0.8		\\
				\hline
			\end{tabular}
			\label{tab:srcsize}
			\end{table}

		\begin{figure}[ht!]
		\centering
		\includegraphics[width=7.5cm]{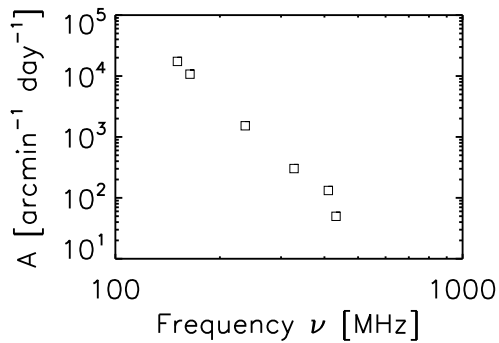}
		\caption{
			Scatter plot of the constants $A$ obtained from fittings similar to that of Figure~\ref{fig:histo_true_src_sizes}. 
			%To remove normalization issues that appear due to the fact that not all six NRH frequencies were used throughly during the solar cycle of interest, 
			%we have derived the normalization constants $A$ taking only events that occurred when all six NRH frequencies were in use.
		}
		\label{fig:histo_true_src_sizes_A}
		\end{figure}
	%Most fitted spectral indices in Figure~\ref{fig:histo_true_src_sizes_2} are close to $\approx$4.	%~\ref{fig:histo_true_src_sizes} and 
	%We have therefore plotted all normalization fitting constants $A_{\nu}$ in Figure~\ref{fig:histo_true_src_sizes_A}.

	As the spectral indices of the distributions for the bursts selected in Section~\ref{sect:selection} were very close (from 4.4 to 4.9) at all frequencies, we decided to plot the normalization constants $A_{\nu}$ as a function of frequency.
	But to remove any normalization issue stemming from the fact that not all frequencies were uniformly employed by NRH during the solar cyle
        (particularly, 150.9 MHz was used only after 2002/10/28), we have used a slightly smaller subset of our data to which we have fitted similar power-laws:
	we have taken only those events that occurred when NRH was observing at all six frequencies simultaneously, and plotted in Figure~\ref{fig:histo_true_src_sizes_A} the normalization 
	constants $A_{\nu}$ of these new  power-law fits.
	Figure~\ref{fig:histo_true_src_sizes_A} suggests a power-law frequency dependence of the normalization constant $A \approx B \, \nu^{\beta}$, which has led us to infer that the distribution for true source FWHM size $s$ could be of the form:
		\begin{equation} \label{eq:size}
			\frac{dN}{ds} = B \, \nu^{\beta} \, s^{\alpha}
		\end{equation}
		We have done two-dimensional fittings of our data, using Equation \ref{eq:size} as model. %WE USED only bins for which counts were above 10.
		The fitting parameters $(B,\alpha,\beta)$ have been found to be moderately dependent on the choice of fitting intervals (and, to a lesser extend, the binning).
		We therefore prefer to give a range of values for which $\chi^2$ near unity was obtained: $(B,\alpha,\beta)$ from (2.0,-3.0,-2.8) to (20.0,-5.0,-3.8), for an average of $\approx$(7,-4,-3.3) over the 2002--2008 period.
		The result is in arcmin$^{-1}$ day$^{-1}$ if $\nu$ is in GHz and $s$ in arcmin.
		To account for the factor $\approx$2 ratio between the 1998--2008 and the 2002--2008 average burst rates (Figure~\ref{fig:histo_frq_time_distribution}), the normalization constant should be changed from 7 to 14 to get a more proper solar-cycle average.

	When the Sun is low on the horizon during certain period of the year, the beam shape can be extremely asymmetric.
	To insure that these extreme cases have not influenced our results, we have therefore run the same study, removing all events occurring in November, December, and January,
	as well as all events within two hours of sunset or sunrise during February and October. The differences with the above results were negligible.

	A more thorough study of source shape and structure, which would address e.g. cases where more than a single source is present \citep[e.g.][and references therein]{Pick1998}, 
	is beyond the scope of this paper.

%+++++++++++++++++++++++++++++++++++++++++++++++++++++++++++++++++++++++++++++++++++++++++++++++++++++++++++++++++++++
\subsection{Radioburst location:}\label{sect:locations}
	
	%In this section, we investigate the location of radio bursts at the six commonly-used NRH frequencies.

		\begin{figure}[ht!]
		\centering
		\includegraphics[width=7.5cm]{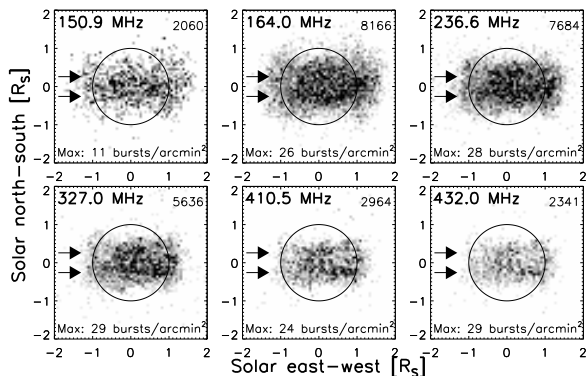}
		\caption{
			Occurrence density distributions for the radio bursts selected in Section~\ref{sect:selection}, on a 1' pixel grid.
			The number in upper right corner is the total number of events with reliable positions.
			The intensity scale is linear from white to black, with black representing the maximum pixel value as specified in each plot.
			The arrows indicate the $\pm$15 degrees heliographic latitudes.
		}
		\label{fig:i_burstdensity_all}
		\end{figure}

		\begin{figure}[ht!]
		\centering
		\includegraphics[width=7.5cm]{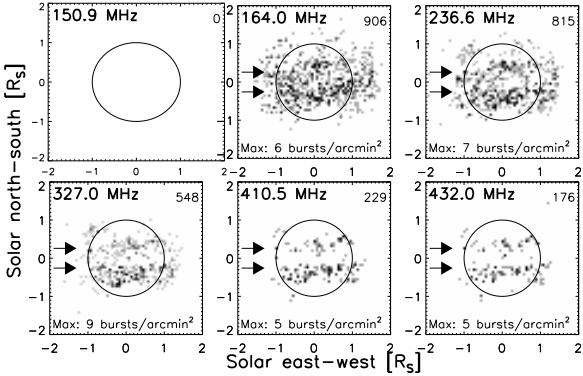}
		\caption{
			As Fig.~\ref{fig:i_burstdensity_all}, taking only events that occurred in 1998. No observations at 150.9 MHz were made that year.
		}
		\label{fig:i_burstdensity_1998}
		\end{figure}

		\begin{figure}[ht!]
		\centering
		\includegraphics[width=7.5cm]{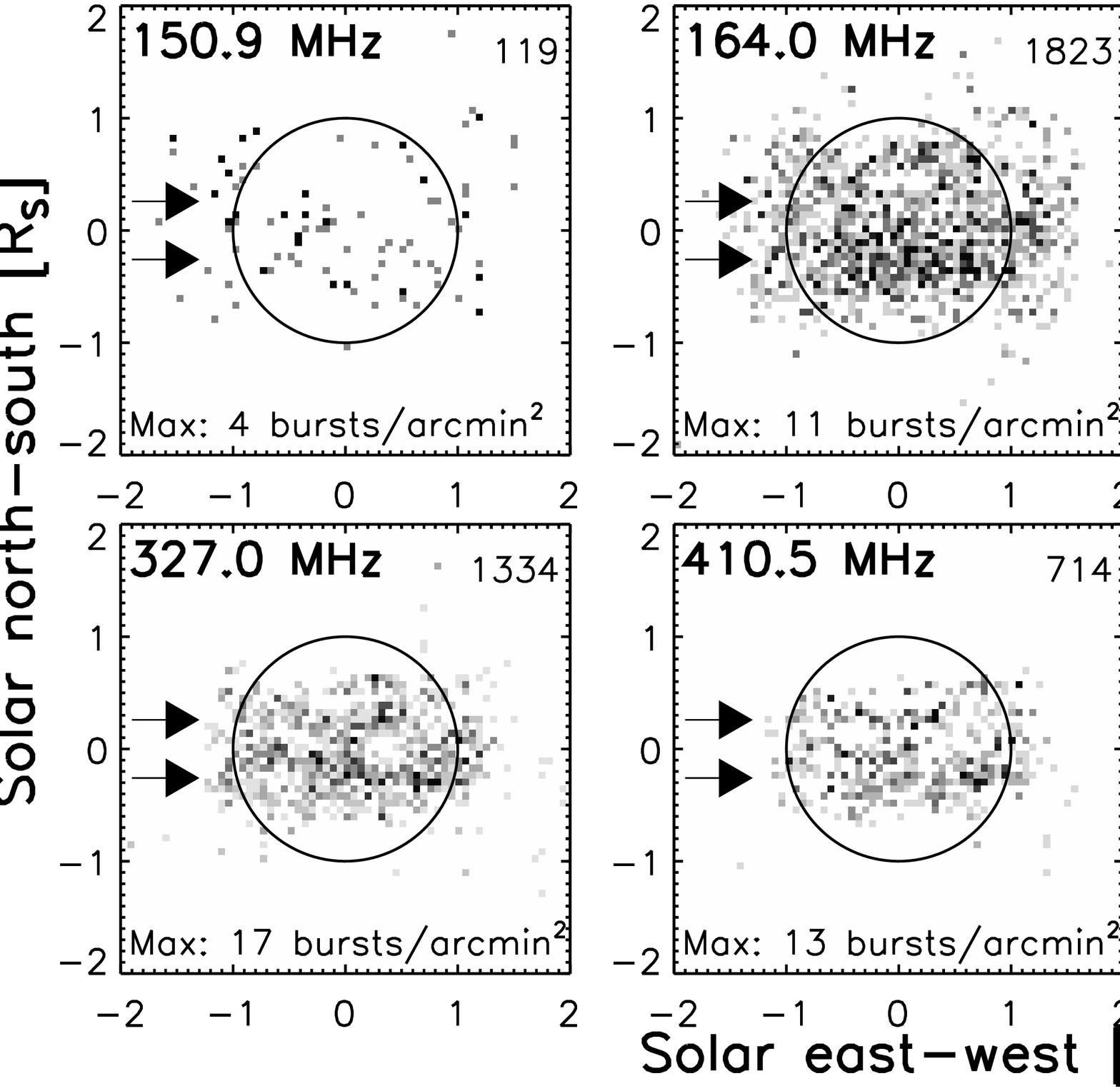}
		\caption{
			As Fig.~\ref{fig:i_burstdensity_all}, taking only events that occurred in 2002.
		}
		\label{fig:i_burstdensity_2002}
		\end{figure}

		\begin{figure}[ht!]
		\centering
		\includegraphics[width=7.5cm]{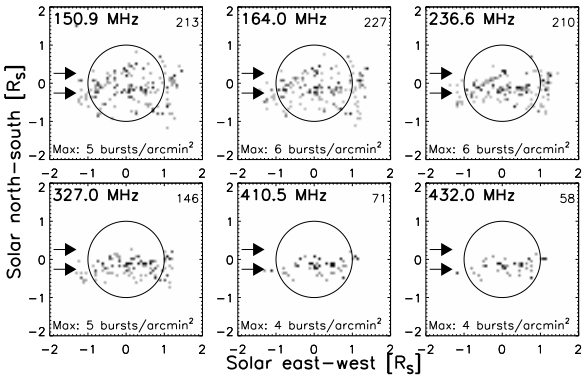}
		\caption{
			As Fig.~\ref{fig:i_burstdensity_all}, taking only events that occurred in 2006.
		}
		\label{fig:i_burstdensity_2006}
		\end{figure}

	%Radioburst positions were determined with two methods: either by taking the brightest pixel in the map, or by taking the gaussian center of the fitted source.
	%Both methods yield very similar results. {\bf I now realize using the fluxoid might have been the best method -- I might do this later on}
	%{\bf Also, the solar P-angle has been corrected for, but not the solar B-angle. Correcting for the B-angle should not lead to big differences, but perhaps to slightly cleaner data.}

	Figure~\ref{fig:i_burstdensity_all} shows the distribution of radio bursts for our selected bursts, using the positions that are likeliest to be correct (as discussed in Section~\ref{sect:selection}),
	while Figures~\ref{fig:i_burstdensity_1998}, \ref{fig:i_burstdensity_2002}, and \ref{fig:i_burstdensity_2006} show the same for three different characteristic years in the solar cycle.
	%The Butterfly diagram in Figure~\ref{fig:butterfly} shows the evolution between 1998 and 2008.

	The most striking feature is, at high frequencies, the concentration of radio bursts in two distinct bands (around latitudes $\pm$15 to $\pm$30 degrees for 432 MHz)
	during early solar maximum years (Figure~\ref{fig:i_burstdensity_1998}), and at lower latitudes as the solar cycle advances.
	This is in remarkable agreement with observed active region and microflare positions and their solar-cycle dependence \citep[e.g.][]{Christe2008b}.
	Later in the solar cycle, bursts are located mostly south of the solar equator, just as active regions were (Higgins et al., 2012, {\it in preparation}).
	
\begin{figure}[ht!]
		\centering
		\includegraphics[width=7cm]{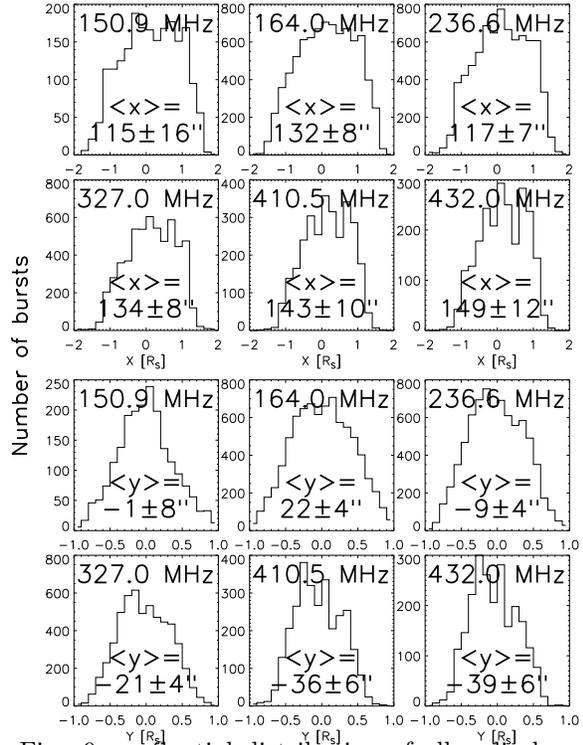}
		\caption{
			Spatial distribution of all radio burst positions between 1998 and 2008: histograms for all six frequencies ({\it top six}: east-west coordinate, {\it bottom six}: north-south coordinate)  with means and their errors.
			A westward shift between 115 and 149'' can be observed at different frequencies.
		}
		\label{fig:histo_loc}
		\end{figure}

	The second most striking feature is the systematic shift of about 2'=120'' westward of the mean position of all radio bursts (Figure~\ref{fig:histo_loc}).
	While a systematic shift of instrumental origin of up to 0.3' westward for all frequencies was expected (from comparison between VLA and NRH positions of a radio spike {\it -- priv. comm. P. Grigis.}),
	clearly it cannot account for all of the observed shift.
	
%+++++++++++++++++++++++++++++++++++++++++++++++++++++++++++++++++++++++++++++++++++++++++++++++++++++++++++++++++++++
\subsection{Source peak brightness temperatures and fluxes} \label{sect:fluxes}

\begin{figure}[ht!]
		\centering
		\includegraphics[width=7.5cm]{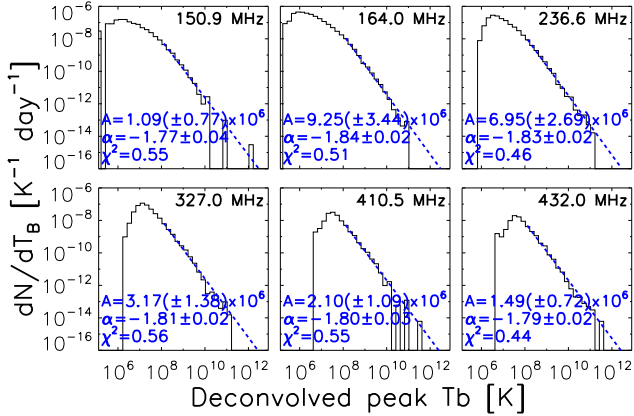}
		\caption{
			Histogram of deconvolved (i.e. assuming both observed and true sources are gaussian-shaped) peak brightness temperature of all Type III bursts.
			%The solid red line is a power-law $f(T_B)=A \, T_B^{\alpha}$ fit to the data, with the best-fit parameters $A$ and $\alpha$ in the upper right corner.
			The dashed blue line is a power-law $\frac{dN}{dT_B}=A \, T_B^{\alpha}$ fit to the data, using the C-statistic \citep{Cash1979}, technically better suited than Poisson statistics for datasets with small number of counts per bin (as is our case for the high-value bins, but in this case leading to negligible differences).
			The associated best-fit parameters are in the lower left corner.
		}
		\label{fig:h_peakintensities}
		\end{figure}

\begin{figure}[ht!]
		\centering
		\includegraphics[width=7.5cm]{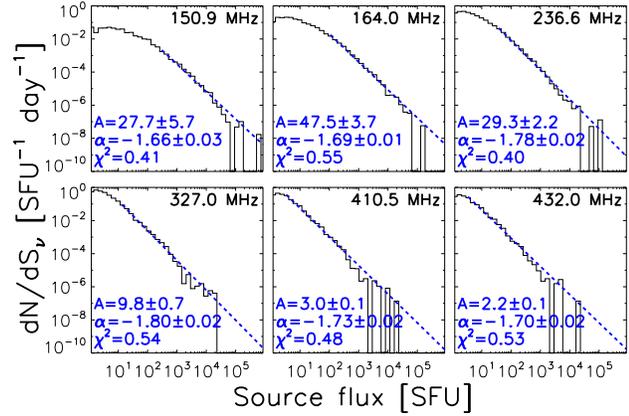}
		\caption{
			Histogram of gaussian source fluxes (averaged over 10 seconds), 
			with power-law fittings $\frac{dN}{dS_{\nu}} = A \, S_{\nu}^{\alpha}$ using the C-statistic \citep{Cash1979}.
		}
		\label{fig:h_fluxes2}
		\end{figure}

\begin{figure}[ht!]
		\centering
		\includegraphics[width=7.5cm]{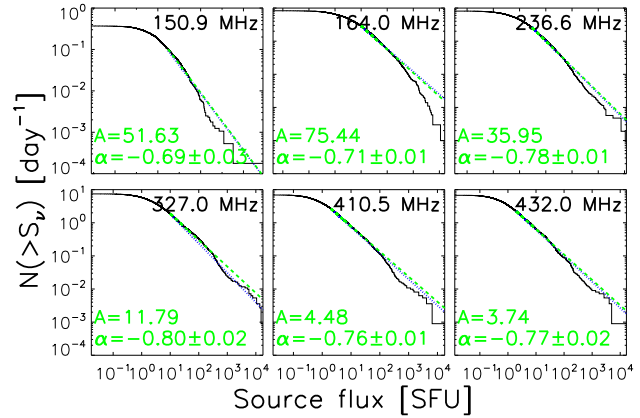}
		\caption{
			%Histogram of gaussian source fluxes (averaged over 10 seconds), 
			%with power-law fittings $\left(\frac{1}{N}\right) \frac{dN}{dS_{\nu}} = \left(\frac{S_{\nu}}{S_{\nu 0}}\right)^{\alpha}$.
			Cumulated data $N(>S_{\nu})$. The blue dotted line is the integral of the fitting found in Figure~\ref{fig:h_fluxes2}, 
			while the green dashed line and coefficients are derived from 
			the Maximum-likelihood method of \citet{Crawford1970}. They match almost exactly.
		}
		\label{fig:h_fluxes2i}
		\end{figure}

	In this section, we investigate the distribution of ``peak'' (in terms of 10-second averages) brightness temperature $T_B$ (in K) and source fluxes $S_{\nu}$ (in sfu, where 1 sfu = 10$^{-22}$ W m$^{-2}$ Hz$^{-1}$).
	Figures~\ref{fig:h_peakintensities} and \ref{fig:h_fluxes2} show that the peak brightness temperatures and fluxes of our radio bursts, 
	follow a $\frac{dN}{dS_{\nu}} \propto S_{\nu}^{-1.7}$ law, very close to what \citet{Nita2002} found.
	Equivalently, the number of bursts above $T_B$ or $S_{\nu}$ follow a power-law with negative index one higher (see Figure~\ref{fig:h_fluxes2i} for $S_{\nu}$). % and \ref{fig:h_peakintensities_i}).
	The roll-over at low fluxes is most likely due to our selection criteria (see discussion in \S\ref{sect:selection}).
	%, while the reason for the paucity of high-flux bursts is unclear. Does it really indicate the absence of such events?

		Note that unresolved sources always have lower brightness temperature than their ``real brightness'' temperature, 
		and that a source's ``real'' brightness temperature is always smaller or equal to its true temperature (opacity).
		For these reasons, the flux $S_{\nu}$ is probably a less misleading quantity to use than $T_B$, and we will concentrate on $S_{\nu}$ in the remainder of this section.

	%Figures~\ref{fig:h_fluxes2_2} and \ref{fig:h_fluxes2i_2} are equivalent to Figures~\ref{fig:h_fluxes2} and \ref{fig:h_fluxes2i}, 
	%except only events occuring during days were all 6 frequencies were used for observations were considered.
	As the spectral indices in Figure~\ref{fig:h_fluxes2} at different frequencies were very close, we decided to plot the normalization constants $A_{\nu}$ as a function of frequency (Figure~\ref{fig:A_NBursts_i_2}).
	As with the source sizes in \S\ref{sect:size_distribution}, we have taken a subset of our data, using only events occurring when NRH was observing with all six frequencies.
	Power-law fittings to these new distributions yield more meaningful normalization constants $A_{\nu}$, plotted in Figure~\ref{fig:A_NBursts_i_2}. 
	%with the following modification: we have taken only events occuring when all six frequencies were used for observations.
	%This was done to remove any normalization issue stemming from the fact that not all frequencies were employed uniformly during the solar cyle
	%(particularly, 150.9 MHz was used only after 2002/10/28, i.e. only later in the solar cycle).	
	Again, notice that the coefficients  $A_{\nu}$ appear to have a power-law dependence with frequency: $A_{\nu} \approx B \, \nu^{\beta}$.
	We have therefore attempted to fit all our data with the following model:

		\begin{figure}[ht!]
		\centering
		\includegraphics[width=7.5cm]{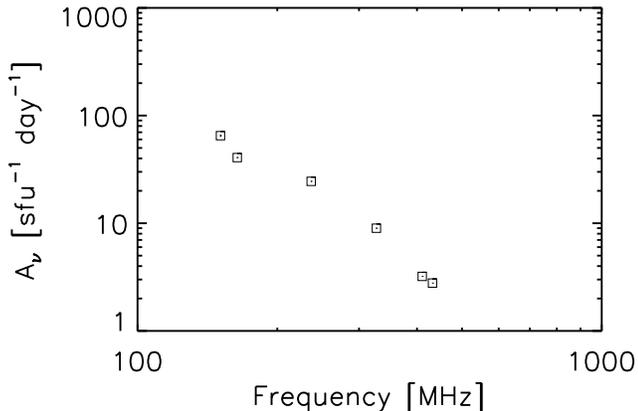}
		\caption{
			Plot of the normalization coefficients $A$ for fittings similar to Figure~\ref{fig:h_fluxes2i}, but taken only over times when all six frequencies were observing, as a function of frequency.
		}
		\label{fig:A_NBursts_i_2}
		\end{figure}

	\begin{equation}\label{eq:flux}
		\frac{dN}{dS_{\nu}} = B \, \nu^{\beta} \, S_{\nu}^{\alpha}
	\end{equation}
	The best-fitting parameters for the 2002--2008 period were found to be (with $\nu$ expressed in GHz, and $S_{\nu}$ in sfu): 
	$B$=0.17$\pm$0.01 sfu$^{-1}$ day$^{-1}$, $\alpha$=-1.7$\pm0.05$, $\beta$=-2.9$\pm$0.1.
	These fitting parameters have proven to be very stable (and all with near unity $\chi^2$) even if changing fitting intervals, and hence much more robust than those found in Section~\ref{sect:size_distribution} for size distributions.
	To account for the factor $\approx$2 ratio between the 1998--2008 and the 2002--2008 average burst rates (Figure~\ref{fig:histo_frq_time_distribution}), the normalization constant should be changed from 0.17 to 0.34 to get a proper solar-cycle average.
	
	%This high $\chi^2$ signifies that the model is not perfect, but gives us a feel for the parameter dependencies.
	%(Taking only bins with more than 10 counts: 0.067054404      -1.3453422      -2.1988870 and 3.4280799.)
	%(Taking only bins with more than 50 counts: 0.23011996      -1.2381623      -1.2761491 and 0.68687628.)
	It is well known that more Type III bursts are seen at low frequencies than at high frequencies, and the above relationship reflects this.
{\bf 	In fact, \citet{Dulk2001} found the spectrum of a Type III burst in the 3--50 MHz range to have a negative spectral index close to 3, in good agreement with our results.
}	

%+++++++++++++++++++++++++++++++++++++++++++++++++++++++++++++++++++++++++++++++++++++++++++++++++++++++++++++++++++++++++++++++++++++++++++++++++++++++++++++++++++++++++++++++++++++++++++++++++
\subsection{Rank-order cross-correlations:} \label{sect:rocorr}

		\begin{figure}[h!]
		\centering
		\includegraphics[width=6.5cm]{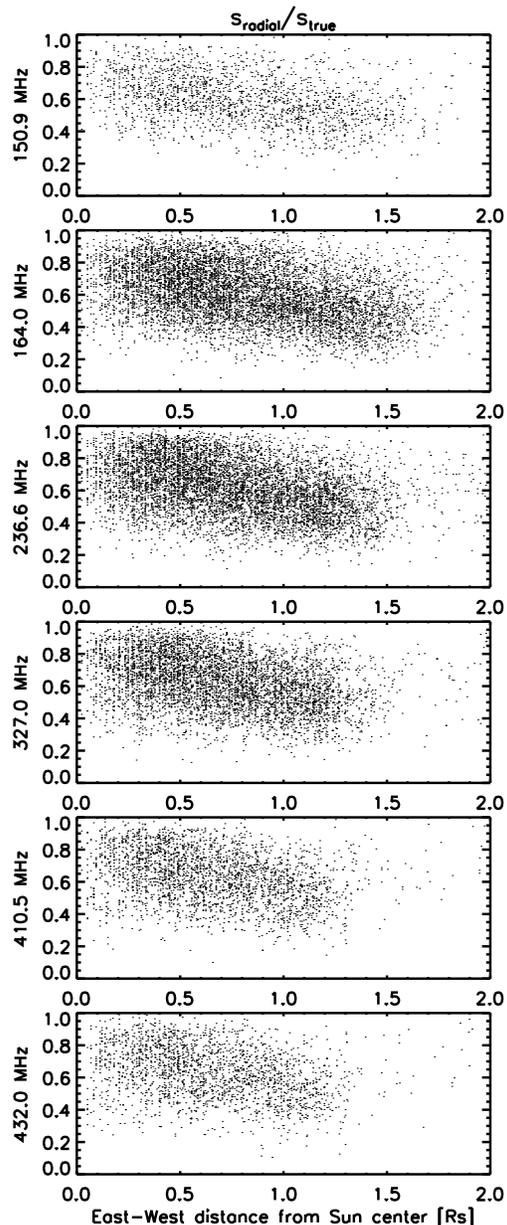}
		\caption{
			Scatter plot of radial distance vs. ``radial elongation'' (ratio of deconvolved source size in the radial direction over deconvolved rms source size) of sources.
			The mild anti-correlation indicates that, the further away from Sun center, the more elongated in the direction perpendicular to the radial sources tend to be.
		}
		\label{fig:r_sradial}
		\end{figure}

	%There are a few weak to moderate correlations, which we may call ``trends'', of note:
	%\begin{enumerate}
	%	\item The size of the sources tend to be larger near the limb, a fact already reported in \citet[e.g.][]{DulkSuzuki1980}, and is thought to be due to scattering and the longer propagation path in the corona.
	%	\item Sources tend to be squashed in the radial direction. This appears like a natural consequence of a source's plasma emission being confined to a narrow altitude range.
	%	\item $(a/b)_{instr}$ and $s_{instr}$ (not shown in Table~\ref{tab:rocorr}) are {\it very strongly} correlated with $|t_{noon}|$, as expected (the beam size and shape is larger and more oval at dawn or dusk). This strong correlation may affect other source shape parameters, particularly if the elliptical gaussian (in either the psf, or for the sources on the Sun) approximation is incorrect.
	%	\item Source fluxes seem uncorrelated with source sizes at low frequency, but a bit more correlated at the higher frequencies. This could be due to the fact that at high frequencies such as 432 MHz, there is a higher chance of having a double source (either because of the better angular resolution, or because of strong refraction effects), and hence more flux appears per unit solid angle.
	%\end{enumerate}

	We cross-correlated most of the observables mined from our study.
	The results and method of correlation are described in Appendix~\ref{appendix:rocorr}.
	The only noteworthy relationship that we have observed is the anti-correlation between the ``radial elongation'' (the ratio of the deconvolved source size in the radial direction over the deconvolved  rms source size) of sources and the radial offset from Sun center (Figure~\ref{fig:r_sradial}).
	This anti-correlation means that sources tend to be circular near disc center, and elliptical near the limb, with minor axis along the radial direction.
	We will discuss the implications of this mild correlation in the next Section.

	There is a noteworthy absence of strong correlation between the source flux $S_{\nu}$ and the source size $s_{true}$, 
	which will also be discussed in Section~\ref{sect:discussion}.

	Due to the longer path in the corona that emissions from sources near the limb have than those near disk center, it is expected that turbulent scattering would make them appear slightly larger on average than their on-disk brethren \citep{Bastian1994}.
	There appears to be indeed a slight correlation ($\approx$0.2) between deconvolved source size $s_{true}$ and angular distance $r$ from Sun center in our data (particularly at the lower frequencies).
	%It appears rather marginal, and hence this fact will only be mentioned here, but not be further discussed.
	
%+++++++++++++++++++++++++++++++++++++++++++++++++++++++++++++++++++++++++++++++++++++++++++++++++++++++++++++++++++++++++++++++++++++++++++++++++++++++++++++++++++++++++++++++++++++++++++++++++++++++++++++++++++++++++++++++++++++++++++++++++++++++++++++++++++++++++++++++++++++++++++++++++
%\subsection{Other things to try, probably in other papers:}

%+++++++++++++++++++++++++++++++++++++++++++++++++++++++++++++++++++++++++++++++++++++++++++++++++++++++++++++++++++++
%+++++++++++++++++++++++++++++++++++++++++++++++++++++++++++++++++++++++++++++++++++++++++++++++++++++++++++++++++++++
%+++++++++++++++++++++++++++++++++++++++++++++++++++++++++++++++++++++++++++++++++++++++++++++++++++++++++++++++++++++
\section{Discussion} \label{sect:discussion}

%\paragraph{General}:
%	About 9000 Type III bursts over $\approx$10 years, with a $\approx$33\% duty cycle implies a bit more than $\approx$8 Type III radio bursts per day (solar cycle-averaged), at least at dm wavelengths.
%	But there are most probably many more events, if one could include those below instrumental limitiations.
%	{\bf Events of a given size or flux occur more often at low frequency than at high frequencies. What to add on this?}

\paragraph{Source sizes:}
	
	The variation of source sizes with frequency relies on a convolution of different effects: 
	the magnetic field structure opening as a function of height in the corona,
	the fact that we may have several different Type IIIs at slightly different locations in the same 10-second time interval, 
	the distribution of intrinsic spatial width of the electron beam \citep[][considered point sources in his work]{Bastian1994},
	and, of course, a host of (mostly refractive in origin) propagation effects \citep[see e.g.][for recent overviews]{PoquerusseMcIntosh1995,Dulk2000}.
	For example, true source sizes could be 3.5 (harmonic) to 1.7 (fundamental) smaller than observed, according to \citet{Melrose1989}.

 	Average values of Type III burst sizes have been measured at frequencies below 169 MHz in the 1970s-1980s \citep{Bougeret1970,Stewart1974,Dulk1979,DulkSuzuki1980}. 
	They are found to increase with decreasing frequencies and, on the average, to be of the order of 5' at 169 MHz 
	\citep[values between 2' and 7', see also][]{Zlobec1992,Mercier2006}, which compares well with the 4.5$\pm$1.6' value in Table~\ref{tab:srcsize}.
	The first spatially resolved observations of Type III bursts at high time resolution with the Nan\c{c}ay Radioheliograph showed that Type III bursts can be resolved into narrow components 
	\citep{RaoultPick1980} (elementary size of the order of 2') with a typical size which increases with time during the burst,
	potentially reaching 5' to 7' at then end of the burst.
	What we have measured in the present paper where we do not use high time resolution data will reflect more the total size of the whole Type III burst group rather than the size of individual components.
	Systematic studies of burst sizes have been performed only at 164 MHz \citep{RaoultPick1980,PickJi1987}.
	The distribution of east-west sizes showed that Type III bursts may be resolved in a few cases in components with real sizes around 2', but that the distribution of sizes can reach 9'.	
	The present study provides the first information on the sizes of Type III bursts at frequencies above 169 MHz.
	It shows that Type III bursts are smaller at high frequencies than at low frequencies and that the size decreases as $\nu^{-3.3}$.
	
	Theory \citep[e.g.][and references therein]{Bastian1994} predicts that scattering of the emission in the solar atmosphere would lead to a $\nu^{-2}$ dependence of the source size.
	This has been confirmed by observations of cosmic sources through the solar corona \citep{Erickson1964,ColesHarmon1989}.
	However, previous observations have showed a quasi-linear dependency of Type III source sizes (from $\approx$0.1 to $\approx$1500 MHz) with wavelength \citep{Steinberg1985,Dulk2000}.
	Our observations support neither.
	
	Although he himself states his analysis becomes invalid for radio wavelengths longer than a few dm, 
	\citet{Bastian1994} predicts a sizeable increase of source sizes near the limb.
	The fact that we seem to observe only little of it might be an indication that there is a cutoff in the level of turbulence.
	I.e. that below a certain altitude, turbulence would no longer be proportional to the ambient density as it is for distances greater than 1.7 $R_S$.
	
\paragraph{Weak correlation between source sizes and fluxes:}
	This is consistent with the interpretation that emission comes from a spatially small region and that propagation effects \citep[see e.g.][for a short overview]{PoquerusseMcIntosh1995} are responsible for the lack of good correlation.
	Notice, however, that the correlation coefficient gradually improves at higher frequencies (from 0.08 at 164 MHz to 0.35 at 432 MHz), suggesting that these (refractive) propagation effects become less important at higher frequencies.

\paragraph{Source anisotropy near the limb:}
	
	The slight anisotropy observed as sources get nearer to the limb can be interpreted as being due to the fact that they are being observed more edge-on than when at Sun center, and that the emitting layer is thinner than the horizontal source size.

\paragraph{Source location:}

	Type III emission at the local plasma frequency (fundamental) is expected to be radiated primarily in a dipolar pattern, perpendicular to the local magnetic field.
	For radiation at the harmonic, a quadrupolar pattern is expected.
	However, the propagation of radio waves is controlled both by large and small scale structures which can modify the primary directivity of the radio emission through respectively refraction \citep[including ducting,][]{Duncan1979} and scattering processes.
	Particularly, emission at the fundamental tends to become aligned with the local density gradient.
	See e.g. \citet{ZheleznyakovZaitsev1970,Melrose1986,Cairns1987a,Cairns1987b,Robinson1994} for theoretical work, \citet{PoquerusseMcIntosh1995} for a recent short overview, and \citet{Thejappa2012} for recent observational work.

	The first investigation of the directivity of Type III bursts was achieved by \citet{CaroubalosSteinberg1974,Caroubalos1974} using stereoscopic measurements at 169 MHz obtained simultaneously with the STEREO-1 experiment and with the NRH. 
	Some directivity was found for the Type III bursts (especially for the fundamental emission). 
	The directivity was however found to be less than for Type I bursts.
	Stereoscopic measurements of the directivity have then been performed using ground-based measurements at 150 MHZ and measurements on ULYSSES in the 1.25--940 kHz range \citep{Poquerusse1996,Hoang1997}.
	They found that the average pattern of the Type III bursts at low frequencies is shifted 40$^{\circ}$ eastward of the radial direction.
	The shift and width is found to decrease with low frequencies. 
	\citet{Bonnin2008} further investigated the directivity of Type III bursts in the interplanetary medium using calibrated  Wind-Ulysses observations in the same frequency range. 
	They confirmed an eastward shift of 23$^{\circ}$ at 940--740 kHz and 55$^{\circ}$ at 55--104 kHz, i.e. increasing with wavelength. 
	In all these papers, the shift is attributed  to a transverse density gradient created by the fast wind (propagating along spiraled open field lines) overtaking the (mostly radial) slow wind.
	At meter wavelengths, an east-west asymmetry for noise storms was reported by \citet{Fokker1960,Fokker1963,Suzuki1961,LeSqueren1963}, 
	\citep[see][pp.111--115, and references within for a review]{Elgaroy1977}. 
	The asymmetry is about 0.1 Rs towards the west. 
	One of the possible explanations which was proposed is that there is a tilt towards the east in the magnetic structures. 

	In the present study performed at decimetric/metric wavelengths, we do not observe a $\propto \lambda$ behaviour of the westward shift in our dataset, 
	but this could be due to the fact that NRH observes bursts at much higher  frequencies and much closer to the Sun.
	On the other hand, combining the observed $\approx$2' westward shift
	with a simple geometric model (details in Appendices~\ref{appendix:tilt}), an average tilt angle between the direction of emission and the radial to Sun center can
	be derived, and has been found to be around 6 degrees.
	(This result is consistent with a more elaborate model using a Monte-Carlo approach, detailed in Appendix~\ref{appendix:tilt2}.)
	This is in the same order of magnitude as was found from noise storms asymmetries.
	If one assumes that the optimal direction of emission {\bf statistically} corresponds to the direction of the local magnetic field,
	one hence concludes that the magnetic field is on average tilted eastward by $\approx$6$^{\circ}$ with respect to the radial direction, at altitudes of a few tenths of a solar radius, where emission at 150--432 MHz occurs.

\paragraph{Derivation of coronal density profile:}
	In the following, we assume that emission is either at the fundamental or the harmonic of the local plasma frequency, though the latter is generally thought to be the dominant emission mechanism in the case of decimetric Type III emission.
	Using position data after 2004/01/25 (the best position data, obtained after the installation of anti-alias antennas), it is possible to derive statistically the 
	average height difference between emission at different frequencies. See Appendix A of \citet{PSH2010} for comprehensive details of the method.
	This method assumes that sources at different frequencies tend to be radially distributed at different heights. 
	The effect of the small $\approx$6$^{\circ}$ tilt angle is neglected.
		\begin{figure}[ht!]
		\centering
		\includegraphics[width=7.5cm]{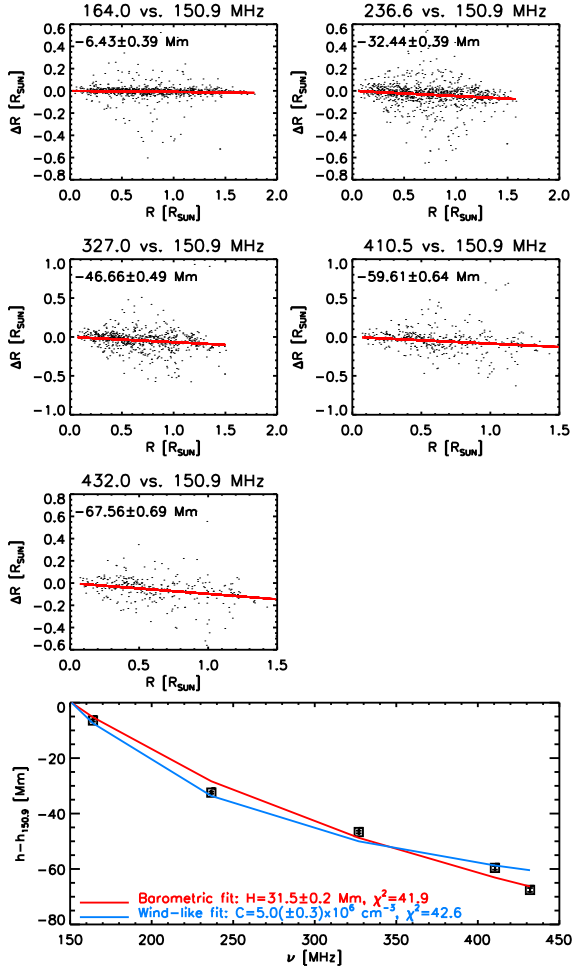}
		\caption{
			Using the methodology presented in detail in Appendix A of \citet{PSH2010}, we have statistically derived the relative heights between emission at different frequencies (first five plots).
			In the bottom plot, we have plotted the height differences, with error bars, and fitted simple density profiles. 
			See text for details.
		}
		\label{fig:density_struct}
		\end{figure}
	In the first five panels of Figure~\ref{fig:density_struct}, we have plotted all average $R$ (plane-of-sky, or POS, distance from Sun center) vs. $\Delta R$ (POS distance between centroid at two different frequencies) for emission at 150.9 MHz versus the other five frequencies.
	The fitted slopes are proportional to the average height difference (the nearer to disk center, the smaller the height difference, and the nearer to the limb; the larger the height difference).
	Note that the line {\it must} go through the origin, making the fittings much stronger than initially appears.
	The height differences are plotted in the last panel of Figure~\ref{fig:density_struct}.
	We have fitted two simple density profiles to this data:
	\begin{enumerate}
		\item A hydrostatic exponential atmospheric $n=n_0 e^{-h/H}$ model (where $H$ is the scale height, and $h$ the altitude above the photosphere), which, in order to be used on our data, is transformed into:
			\begin{equation}
				\Delta h_{ij} = h_i-h_j = -2H \ln \frac{\nu_i}{\nu_j}
			\end{equation}
			Where $i$,$j$ refer to different frequencies of observations.
			Note that this model is not influenced by whether the emission is at the fundamental or harmonic of the local plasma frequency.
			Best fitting yields $H$=31.5 Mm (the red curve in the bottom plot of Figure~\ref{fig:density_struct}).
			This scale height corresponds to a $\approx$0.6 MK corona, lower than the more usual quiet Sun values of 1--1.5 MK.
			This could be due to the fact that we are observing Type IIIs near regions of open field lines, and/or the fact that we are near altitudes where solar wind-like conditions start to take over.
			\citet{David1998} indeed found very low temperatures at similar heights in polar coronal holes.
				
		\item A ``solar wind-like'' atmosphere $n(h)=C (h/R_S)^{-2}$ model \citep{Cairns2009}, which, in order to be used on our data, is transformed into:
			\begin{equation}
				\Delta h_{ij} = h_i-h_j = 8980 \, m \, R_S \,\sqrt{C} \left( \frac{1}{\nu_i} - \frac{1}{\nu_j} \right)
			\end{equation}
			Where $m$=1 or 2 for emission either at fundamental of harmonic.
			The normalization constant $C$ can be viewed as the electron density at an altitude of 1 $R_S$.
			Best fitting yields $C$=5$\times$10$^{6}$ cm$^{-3}$ (emission at fundamental) and $C$=1.2$\times$10$^{6}$ cm$^{-3}$ (emission at harmonic), and the blue curve in the bottom plot of Figure~\ref{fig:density_struct}.
			The numbers for harmonic emission are comparatively low, while for fundamental emission, they are comparable to the Baumbach--Allen formula and other published work: e.g. \citet{Newkirk1961,Mann1999} for theoretical models, 
			and see \citet{FainbergStone1974} and \citet{AschwandenActon2001} for a recent compilation of observational work.

	%	\item A \citet{Mann2005} atmosphere (their Equation 3), which, in order to be used on our data, can be transformed into:
	%		\begin{equation}
	%			h_i-h_j = R_S \left[ \frac{1}{1+ \frac{1}{D} \ln(\frac{n_i}{n_s})} - \frac{1}{1+ \frac{1}{D} \ln(\frac{n_j}{n_s})} \right]
	%		\end{equation}
	%		With $n_i=(\frac{\nu_i}{8980})^2$, $D=31.83/T$.
	%		Best fitting yields $T$=4.6 MK, $n_s$=3.6x10$^6$ cm$^{-3}$, and the purple curve in the bottom plot of Figure~\ref{fig:density_struct}.
	\end{enumerate}
	%For comparison, we have also added the density model from \citet{Mann2005} (their equation 3) (green curve), with $T$=0.6 MK, $n_s$=9x10$9$ cm$^{-3}$.
	%Conclusion: 
	
	Note that we have also used the pre-2004 data, and have found there was very little difference with the ``better'' post-2004 data.
	Both fittings appear equally good (or equally bad), and this methodology cannot decisively choose which of these two models is the most appropriate.	

\paragraph{Source intensities and fluxes:}

	Maximum observed brightness temperature $T_B$ in interplanetary Type III bursts are reported to be $\approx$10$^{15}$ K, and up to $\approx$10$^{12}$ K for coronal Type III bursts \citep{Melrose1989}.
	We observe a few events $>$10$^{11}$ K in our dataset.
	%Note that unresolved sources always have lower brightness temperature than their real brightness temperature (filling factor), 
	%and that a source's brightness temperature is always smaller or equal to its true temperature (opacity).
	%For these reasons, we generally prefer to use $S_{\nu}$ than $T_B$, as it is probably more accurate.

	We have found a power-law dependence of the occurrence rate of Type III bursts of peak flux $S_{\nu}$ with a negative spectral index of 1.7,
	very similar to \citet{Nita2002}, and to what is found in X-rays for flare energies \citep{Crosby1993,Hannah2008}.
	\citet{Eastwood2010} have found a 2.1 index for interplanetary Type III solar radio storm in the 0.125-16 MHz range..
	A 1.7 negative power-law index is also an oft-seen value in self-organized criticalitity (SOC) studies: \citet{Bak1988,LuHamilton1991, Vlahos1995, Georgoulis1996, Aschwanden2012}.
	%{\bf Does Wheatland have something relevant...?}.
	In their study of Type I noise storms, \citet{Mercier1997} had found negative power-law spectral indices of about 3 to 3.5 for their $\frac{dN}{dS_{\nu}}$,
	and have attributed it to the predicted low-energy part of SOC theory \citep{Vlahos1995}.
	(Type I noise storm bursts are indeed less energetic but much more frequent than the decimetric Type III bursts studied here.)
	Similarly, \citet{Morioka2007} found a -3.6 index for interplanetary ($\approx$MHz) ``micro-type-III'' bursts.

	Additionally, we find a $\propto \nu^{\beta}$ (with $\beta \approx$2.9) spectral dependence of $\frac{dN}{dS_{\nu}}$, reflecting the fact that Type IIIs are more easily observable, or occur more often, at 164 MHz than at 432 MHz.
	This exponent could provide constraints to Type III emission models such as in \citet{RobinsonCairns1998, Melrose1989}.
	
	%Figure 1 of \citet{Melrose1989} seem to show that coronal bursts have a $T_B \propto \nu^{-8}$ dependence...?
	%In any case, $T_B$ can be far from the truth for unresolved sources, and thus we prefer to stick to $S_{\nu}$.

\paragraph{Solar-cycle averaged radiative output:} 
		
		\begin{figure}[h]
		\centering
		\includegraphics[width=7.5cm]{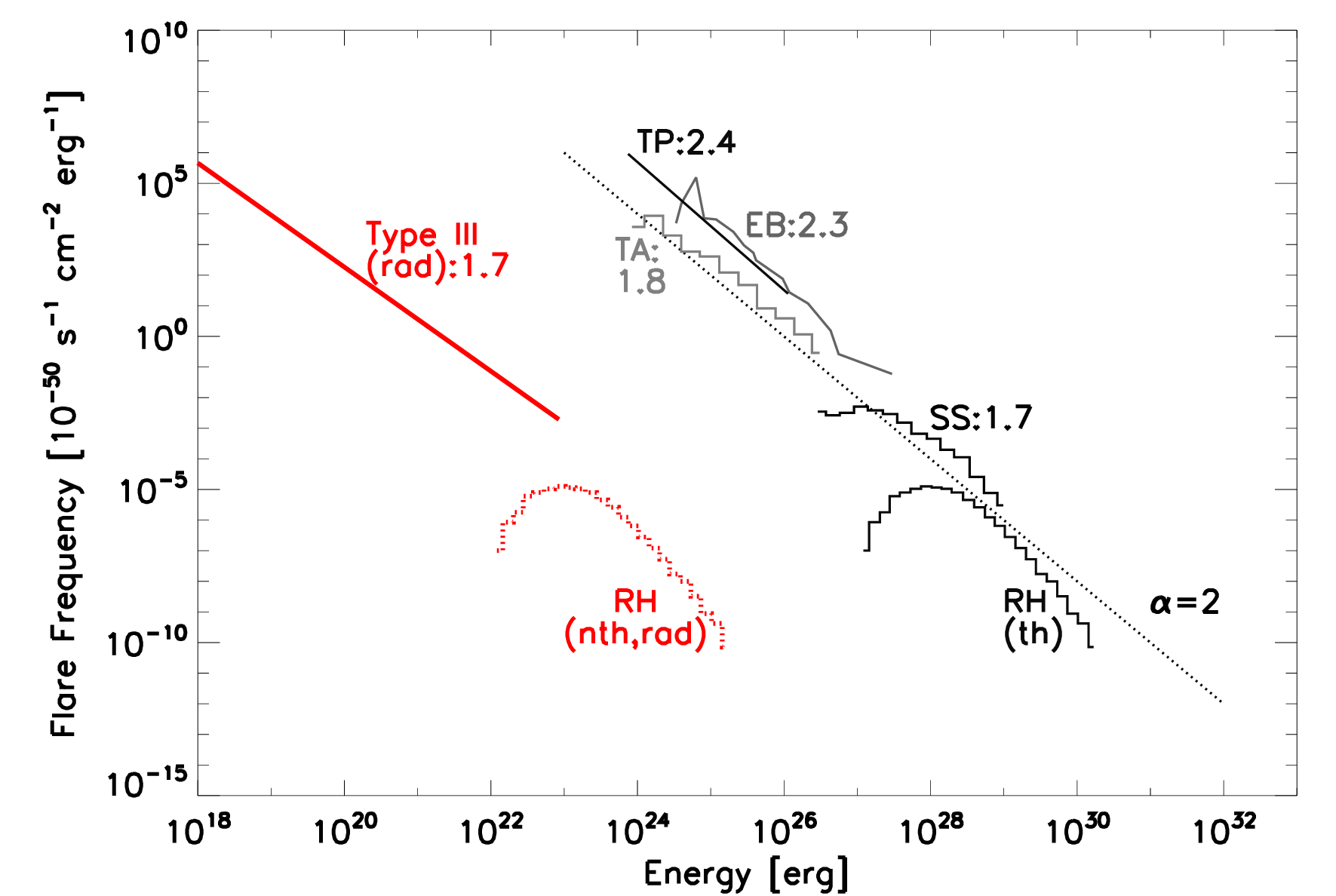}
		\caption{
			%{\it Solid line}: Flare heating input to the solar corona as derived from various EUV/X-ray microflare studies \citep[dotted line in][]{Hannah2008}.
			%{\it Dashed line}: energy {\it radiated} by our Type IIIs (see text for details).
			{\it Black/gray:} Approximate flare heating input to the solar corona as derived from various EUV/X-ray nanoflare and microflare studies:
			{\it RH: RHESSI}, \citet{Hannah2008}; {\it SS: Yohkoh} SXT, \citet{Shimizu1995}; {\it TA: TRACE}, \citet{Aschwanden2000}; {\it TP: TRACE}, \citet{ParnellJupp2000};
			{\it EB: SOHO EIT}, \citet{BenzKrucker2002} \citep[after][]{Hannah2008}.
			{\it Red lines:} Radiated energies. 
			The dotted red line corresponds to the radiated energy in non-thermal HXR that roughly accompanied the RHESSI-derived flare input,
			using an approximated nonthermal bremsstrahlung efficiency of 10$^{-5}$.
			The solid red line is the radiated energy derived from this Type III study.
		}
		\label{fig:dm_burst_freq}
		\end{figure}

		To put things in perspective, we compared the global energy radiated by the Type III bursts and the energy input of flares into the corona.
		We have plotted in Figure~\ref{fig:dm_burst_freq} the {\bf ``Radio Burst Occurrence Rate'' (RBOR)} of our observed Type IIIs and compared it to the energy input from EUV/X-ray events to the corona, derived in other studies
		\citep[see e.g.][for a discussion on this topic]{Hudson1991}.
		The RBOR can be estimated from the power-law fitting parameters by using the $\frac{dN}{dS_{\nu}}$ at 164 MHz (Figure~\ref{fig:h_fluxes2}), % Eq.~\ref{eq:flux},  %$\nu$=164 MHz from Figure~\ref{fig:h_fluxes2i}, 
		by assuming a burst bandwidth of $\Delta \nu \approx 300$ MHz \citep{IslikerBenz1994c}, 
		taking $\Delta t$=10 seconds (since all our numbers are 10-second averages),
		and computing the total energy radiated as observed from $D$=1 AU.
		We have further normalized to the photospheric surface area, in order to compare to the ``Flare Frequency'' often used in coronal heating studies:
		\begin{equation}
			RBOR = \frac{1}{4\pi R_S^2} \frac{1}{4\pi D^2 \Delta\nu \Delta t}\frac{dN}{dS_{\nu}}
		\end{equation}
		(Note that $4\pi D^2 \Delta \nu \Delta t dS_{\nu} = dE$ is the radiated energy for one event).
		The RBOR is hence the occurrence rate of radiated energy distribution of Type III events, per unit (photospheric) solar area.
		%Note that the strongest Type IIIs ($\approx$2$\times$10$^4$ sfu) radiate $\approx$10$^{21}$ ergs, only a few orders of magnitude less than an EUV nanoflare.
		One notices that the {\it radiated} energy by decimetric Type IIIs is about five to six orders of magnitude less than the energy input by EUV nanoflares in the corona
		(we loosely refer to flares observed in EUV only as ``nanoflares'', and those observed also in X-rays as ``microflares'' or ``flares''),
		about the same ratio that exists between the non-thermal hard X-ray (HXR) emission associated with X-ray flares and microflares, though the emission mechanism is believed to be vastly different (coherent plasma emission vs. bremsstrahlung).
		%It is interesting to note that 10$^{-7}$-10$^{-5}$ is the ratio of the typical non-thermal bremsstrahlung efficiency in the case of the thick-target model (varying with injected electron spectral index and low-energy cutoff, see Appendix~\ref{appendix:brmeff}),
		%i.e. the HXR radiated power distribution from flares and microflares is close to that of coronal Type III bursts.		
		\citet{Ramesh2010} have estimated that a 40 sfu metric (77 MHz) interplanetary Type III burst was produced by a $\approx3\times10^{24}$	ergs electron beam.
		Using the same assumptions on $\Delta\nu$, $\Delta t$, and $D$ as before, this would imply an efficiency of non-thermal radiated energy over beam energy of $\approx$10$^{-5}$ (within an order of magnitude).
		It is interesting to note that this ratio is somewhat similar to the one between radiated non-thermal X-rays and the energy in thick-target \citep{Brown1971} accelerated electrons (see Appendix~\ref{appendix:brmeff}).
		To our present knowledge, there is no obvious reason for the non-thermal bremsstrahlung efficiency (an incoherent process) and the efficiency of Type III plasma emission (a collective process) to be similar, and it may be a coincidence.
		Clearly, more studies are needed to acertain radiative efficiencies, but assuming they are indeed similar, this has for consequence that nanoflare and possibly microflare electrons could be similar in number and energies as Type III-producing escaping coronal electron beams!
		This is in contradiction with recent studies \citep{Christe2008,PSH2009a}, which have shown that there is probably a factor $\approx$500 more flare electrons than escaping electrons.
		However, note that these studies were centered on one-to-one association of precipitating flare electron beams and escaping interplanetary electron beams (loosely assuming that interplanetary beams and their counterparts in the low corona have similar properties), 
		while the present study is more global (and, for example, it allows for temporal displacement between the two phenomena and different rates of occurrence).
		One of the implications is that escaping electron beams carry energy away from the corona, partially or wholly negating coronal heating by nanoflares.
		We dare not speculate any further without a better handle on radiative efficiencies of Type III-producing electron beams.
		The upcoming NuStar mission \citep{Harrison2005}, and perhaps also the FOXSI rocket flight, should reveal how many electrons are in Type III beams (via their X-ray emission), yielding the efficiency of Type III radio emission.

\section{Summary, conclusions, and future work} \label{sect:ccl}
	
	We have studied solar Type III radio bursts observed by NRH during 1998--2008, almost a full solar cycle.
	From this very rich dataset, we have determined source sizes, locations, fluxes, and their frequency distributions.

	The source size distribution $\frac{dN}{ds}$can be fitted with a double power-law in $\nu$ and in $s$.
	The Type III burst sizes are found to decrease as $\nu^{-3.3}$, which could reflect a combination of the magnetic field opening 
	as a function of height in the corona, and the distribution of the spatial widths of electron beams in the corona.

	The source flux distribution follows a -1.7 power-law at all frequencies. 
	A two-dimensional power-law fitting yields $\frac{dN}{dS_{\nu}} \propto \nu^{-2.9} \, S_{\nu}^{-1.7}$.
	These values offer additional constraints to theoretical models of Type III emission.
	
	Type III emission generally come from active regions (just as flares), and we have observed an east--west asymmetry in their location, 
	which could be explained by an eastward tilt of the magnetic field (compared to the local radial to suncenter) at the typical height where our metric/decimetric Type III emission occurred.
	We have estimated that the tilt angle is about $\approx$6$^{\circ}$ at altitudes $\approx$0.3--0.6 $R_S$. 

	A barometric fit to the data led to a $\sim$32 Mm scale height, and temperatures similar to what can be found in coronal holes, 
	while fitting a solar-wind like density profile yielded a density of $\frac{5 \times 10^{6}}{m^2}$ cm$^{-3}$ (where $m$ is the harmonic number) at an altitude of 1 $R_S$.

	Furthermore, we speculated that the solar-cycle averaged radiated Type III energy output could be similar to that of radiated non-thermal bremmstrahlung from accelerated nanoflare electrons.
	%Although further work on the topic is definitely needed, it is hinted that both radiative efficiencies could be similar within an order of magnitude.
	We further suggest that escaping electron beams could be a viable mechanism for carrying energy away from the corona, in a quantity similar to the energy introduced into it through nanoflares,
	though this result depends heavily on as yet ill-known radio radiative efficiencies.

	It is likely that statistical testing of the different theoretical models of the generation and propagation of Type III bursts (e.g. \citeauthor{Dulk1979},\citeyear{Dulk1979}; \citeauthor{Duncan1979},\citeyear{Duncan1979}; ...; \citeauthor{RobinsonCairns1998},\citeyear{RobinsonCairns1998}) could be done.
	In particular refractive effects on apparent source sizes, locations, and fluxes.
	Possible other future work includes the usage of sub-second or the full 128 ms resolution data, if it ever becomes easily usable (e.g. stored on a file system instead of on tapes).
	After the development of a good (and automated) Type III burst detection and discrimination \citep[e.g.][]{Lobzin2009}, this work could be repeated almost verbatim. %, with slightly more accurate occurrence rates.
	In the meantime, it is envisaged to do a statistical comparison between the Type III bursts in this work and the X-ray flare parameters found in the RHESSI flare or microflare lists,
	and get observational constraints on radiative efficiencies by comparing with in-situ measurements of electron beams associated with the metric/decimetric Type III emissions presented in this work.

	%This dataset could be further used to study other depencies, such as source fluxes vs. radial offset, diurnal effects due to Earth atmospheric refraction, source aspect ratio, correlations with other wavelengths (e.g. RHESSI microflares).

%==========================================================================================================================================================================================	
%==========================================================================================================================================================================================	
%==========================================================================================================================================================================================	
\appendix

\section{Convolution of two elliptical gaussians with different tilt angles.} \label{appendix:deconvolution}

	The convolution of a gaussian elliptical source (semi-major axis $a_{true}$, semi-minor axis $b_{true}$, and tilt angle $\theta_{true}$) 
	with a gaussian elliptical beam ($a_{beam}$, $b_{beam}$, $\theta_{beam}$)
	yields an observed source that is also an elliptical gaussian ($a_{obs}$, $b_{obs}$, $\theta_{obs}$).
	Through Fourier transformation, the following can be derived:
	\begin{eqnarray}
		\tan (2 \, \theta_{true}) & = & \frac{ (a_{obs}^2 - b_{obs}^2) \, \sin (2 \, \theta_{obs})   -  (a_{beam}^2 - b_{beam}^2) \, \sin (2 \, \theta_{beam}) }{ (a_{obs}^2 - b_{obs}^2) \, \cos (2 \, \theta_{obs})   -  (a_{beam}^2 - b_{beam}^2) \, \cos (2 \, \theta_{beam}) } \\
		a_{true} & = & \sqrt{ \frac{1}{2} \left( a_{obs}^2 + b_{obs}^2 - a_{beam}^2 - b_{beam}^2 + \Delta \right) }	\\
		b_{true} & = & \sqrt{ \frac{1}{2} \left( a_{obs}^2 + b_{obs}^2 - a_{beam}^2 - b_{beam}^2 - \Delta \right) }
	\end{eqnarray}
	with:
	\begin{eqnarray}
		&&	\Delta = \frac{ (a_{obs}^2 - b_{obs}^2)\,\cos(2\,\theta_{obs}) - (a_{beam}^2 - b_{beam}^2)\,\cos(2\,\theta_{beam}) }{ \cos(2\,\theta_{true})} \\
		&or&	\Delta = \frac{ (a_{obs}^2 - b_{obs}^2)\,\sin(2\,\theta_{obs}) - (a_{beam}^2 - b_{beam}^2)\,\sin(2\,\theta_{beam}) }{ \sin(2\,\theta_{true})}		
	\end{eqnarray}
	(use whichever has non-zero denominator).

	Moreover, an important result, which can be derived by combining Eqs (A2) and (A3):
	\begin{equation}
		s_{obs}^2 = s_{true}^2 + s_{beam}^2
	\end{equation}
	where $s_i^2 = a_i^2 + b_i^2$, with $i$ being $obs$, $true$, or $beam$.
	I.e., when dealing with the convolution of elliptical gaussians, the rms size of the observed source is simply the quadratic mean of the rms size of the beam and of the true source,
	independent of any tilt angle between the true source and the beam.

\section{Non-thermal bremsstrahlung efficiency in the case of the thick-target model:} \label{appendix:brmeff}
	For an injected electron spectrum with negative spectral index $\delta$ and low-energy cutoff $E_c$, and assuming non-relativistic cross-sections for bremsstrahlung (Bethe-Heitler) and energy losses,
	the ratio of total emitted X-ray energy over accelerated electron energy can be approximated by the following analytical expression, derived by considering the photon spectrum to be a power-law above a low-energy cutoff $E_c$, and a flat spectrum below it:
	\begin{equation}
		\eta \approx \frac{4}{3\pi}\frac{\alpha}{\Lambda} \overline{z^2} \frac{1}{m_ec^2} \frac{B(\delta/2,1/2)}{\delta-1} f^{-\delta+2} \left( 1 + \frac{f E_c}{\delta-3} \right)
	\end{equation}
	where $\alpha$ is the fine structure constant ($\approx$ 1/137), $\Lambda$ is the Coulomb logarithm ($\approx$20-30 for fully ionized coronal plasma), $\overline{z^2}$ is the average coronal atomic number-squared ($\approx$1.44), 
	$m_ec^2$=511 keV, $B$ is the Beta function, $\delta$ is the electron flux power-law negative spectral index, $E_c$ is the low-energy cutoff expressed in keV, and $f$ is a slowly-varying function that depends mostly on $\delta$, and is equal to about 0.3 for $\delta$=4.
	For example, for $\delta$=4 and $E_c$=10 keV, one gets a bremsstrahlung emission efficiency of 8.5$\times$10$^{-6}$.
	We have found good (within a 10--20\%) correspondence with (exact) numerical computations.

\section{Magnetic field tilt estimation through simple geometry arguments} \label{appendix:tilt}

	From the observation that radio sources seem to be displaced westward by $\sim$2' on the average, we derive an average tilt angle of the magnetic field (or at least the direction of the main Type III emission) with the local radial to Sun center.

\begin{wrapfigure}{l}{0.31\textwidth}
  \vspace{-20pt}
	\begin{center}
	\includegraphics[width=6cm]{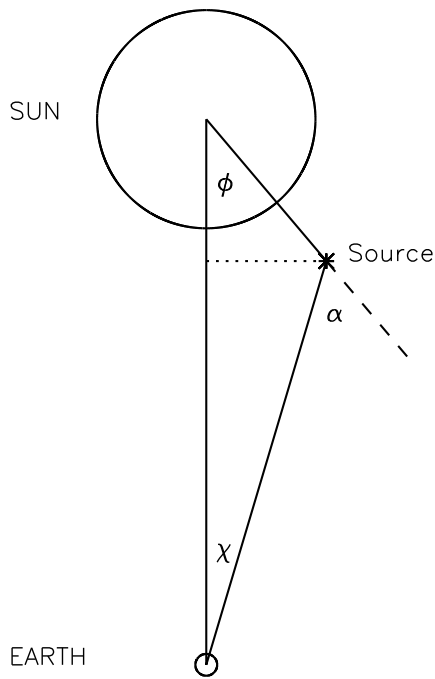}
	\end{center}
  \vspace{-20pt}
	\caption{Type III emission geometry.}\label{fig:T3emissiongeometry}
  \vspace{-0pt}
\end{wrapfigure}

	For simplicity, we consider in the following the geometry to be wholly within the ecliptic plane,
	and the Type III emission to have infinite directivity (zero beamwidth).

	The ``average'' source is observed to be at a westward angle $\chi$ from Sun center ($\chi \approx$ 2'), at an altitude $R_S+h$, where $R_S$ is the solar radius.
	We call $\phi$ the angle between the Sun-Earth line and the radial to the source, and $\alpha$ the angle between the radial to the Sun and the direction in which the Type III is emitted (Figure~\ref{fig:T3emissiongeometry}).
	%I.e. $\alpha$ is the angle, at the source, between the radial to the Sun and the source-Earth line.
	Then, the following must be true for the emission from the ``average'' source (which is near the solar surface) to be observed at Earth:

	\begin{equation}\label{eq:appendix:tilt:1}
		\alpha = \phi + \chi
	\end{equation}
		
	A bit of trigonometry yields:
	\begin{equation} \label{eq:appendix:tilt:2}
		\sin(\phi) = \frac{ D - (R_S+h) \cos(\phi)}{R_S+h} \tan(\chi)
	\end{equation}
	where $D$ is the Sun-Earth distance ($\approx$215 $R_S$).
	This is a transcendental equation which can be iteratively solved for $\phi$.
	$\alpha$ is then simply derived from Eq~\ref{eq:appendix:tilt:1}.
	The following explicit formula, derived using simplified geometry,
	\begin{equation} \label{eq:appendix:tilt:3}
		\alpha \approx \chi + \arcsin\left(\frac{\chi \cdot D}{R_S+h}\right)
	\end{equation}
	has accuracy better than 1\% for $h \leq 0.6 R_S$.

	Using $\chi$=2', and for $h$ between 0.1 and 0.5 $R_S$ (approximate altitude up to which decimetric emission can be found, cf. Figure~\ref{fig:histo_loc}), one obtains $\alpha$ between 6.6 and 4.8 degrees. 
	Say $\alpha$=6$\pm$1 degrees.
	
	These results were derived from a very simple 2D model, assuming the Type III emission had infinite directivity.
	but they are in good agreement with the Monte-Carlo approach presented in Appendix~\ref{appendix:tilt2}.

\section{Magnetic field tilt estimation through a simple Monte-Carlo approach} \label{appendix:tilt2}

\begin{figure*}[h]
		\centering
		\includegraphics[width=15cm]{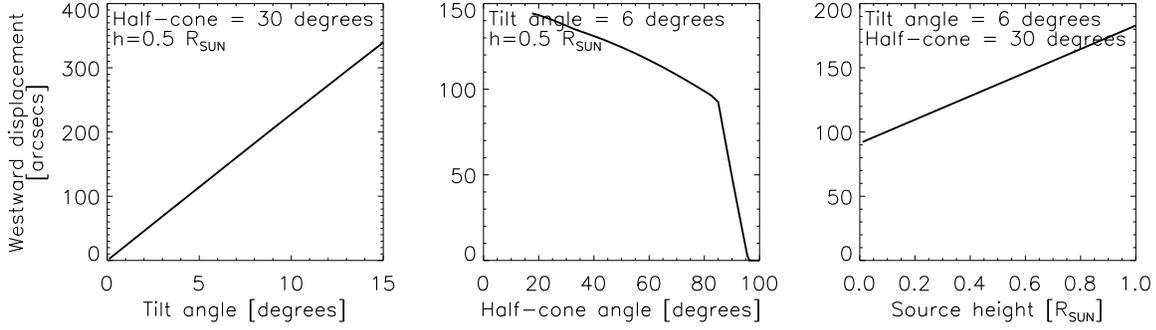}
		\caption{
			Westward displacement of the average source position, obtained through a Monte-Carlo simulation.	
			See text for details.			
		}
		\label{fig:sim_avg_xpos}
		\end{figure*}

	In addition to the simple geometrical approach described in Appendix~\ref{appendix:tilt}, 
	we ran a simple 3D simulation to estimate the effects of a tilt angle $\alpha$ of the radio emission with respect to the local radial,
	the half-width $\Delta\alpha$ of the cone of emission, and the source altitude $h$.
	A hundred thousand sources were uniformally distributed in longitude, but only within 15 and 30 degrees in latitude (which reflects reality).
	For each source, if Earth (and the observer) was within the cone of emission, the apparent position of the source on the Sun was recorded.
	If not, it was omitted.
	In Figure~\ref{fig:sim_avg_xpos}, we display the variations in averaged source East-West position, if one varies $\alpha$, $\Delta\alpha$, or $h$.

	Using canonical values of $\alpha$=6 degrees, $\Delta\alpha$=30 degrees, and $h$=0.5 $R_S$, one obtains a westard displacement of 137'', i.e. very close to our reported observations.
	It can be said from Figure~\ref{fig:sim_avg_xpos} that the result does not strongly depend on the exact value of $h$, i.e. a change of 0.1 $R_S$ leads to a change of the average displacement by $\approx$10''.
	Reasonable ($<$80 degrees) values of $\Delta\alpha$ lead to variations of up to $\pm$25'' at the most.

	We conclude that the 6$\pm$1 degrees tilt angle obtained in Appendix~\ref{appendix:tilt} is further validated by the simple 3-D model presented in this appendix.

\section{Rank-order correlations} \label{appendix:rocorr}

\begin{table*}[ht!]
\caption{164, 327, \& 432 MHz rank-order cross-correlations between parameters.}
\rotatebox{90}{
\begin{minipage}[b][8cm]{18cm}
%\begin{tiny}
\centering
\begin{tabular}{l|ccccccccccc}
\hline
 & \rotatebox{90}{$r$} & \rotatebox{90}{$T_{B,true}$} & \rotatebox{90}{$S_{\nu}$} & \rotatebox{90}{$s_{true}$} & \rotatebox{90}{$\theta_{true}$} & \rotatebox{90}{$(a/b)_{true}$} & \rotatebox{90}{$\frac{s_{true,rad}}{s_{true}}$} & \rotatebox{90}{$\frac{s_{true,azim}}{s_{true}}$} \\
\hline\hline
164 MHz \\
\hline
$r$ & 1.00 & -0.16 & -0.07 & 0.28 & 0.00 & 0.18 & -0.37 & 0.37 \\
$T_{B,true}$ & -0.16 & 1.00 & 0.90 & -0.16 & -0.00 & -0.01 & 0.13 & -0.02 \\
$S_{\nu}$ & -0.07 & 0.90 & 1.00 & 0.08 & 0.02 & 0.01 & 0.01 & -0.01 \\
$s_{true}$ & 0.28 & -0.16 & 0.08 & 1.00 & 0.07 & 0.31 & -0.13 & 0.24 \\
$\theta_{true}$ & 0.00 & -0.00 & 0.02 & 0.07 & 1.00 & 0.06 & 0.01 & -0.02 \\
$(a/b)_{true}$ & 0.18 & -0.01 & 0.01 & 0.31 & 0.06 & 1.00 & -0.34 & 0.45 \\
$s_{true,radial}/s_{true}$ & -0.37 & 0.13 & 0.01 & -0.13 & 0.01 & -0.34 & 1.00 & -0.89 \\
$s_{true,azimuthal}/s_{true}$ & 0.37 & -0.02 & -0.01 & 0.24 & -0.02 & 0.45 & -0.89 & 1.00 \\
\hline\hline
327 MHz \\
\hline
$r$ & 1.00 & -0.06 & 0.07 & 0.21 & 0.01 & 0.05 & -0.34 & 0.34 \\
$T_{B,true}$ & -0.06 & 1.00 & 0.77 & -0.21 & -0.00 & -0.06 & 0.11 & 0.00 \\
$S_{\nu}$ & 0.07 & 0.77 & 1.00 & 0.29 & 0.04 & 0.01 & -0.01 & 0.01 \\
$s_{true}$ & 0.21 & -0.21 & 0.29 & 1.00 & 0.07 & 0.25 & -0.05 & 0.16 \\
$\theta_{true}$ & 0.01 & -0.00 & 0.04 & 0.07 & 1.00 & -0.00 & -0.01 & -0.00 \\
$(a/b)_{true}$ & 0.05 & -0.06 & 0.01 & 0.25 & -0.00 & 1.00 & -0.22 & 0.34 \\
$s_{true,radial}/s_{true}$ & -0.34 & 0.11 & -0.01 & -0.05 & -0.01 & -0.22 & 1.00 & -0.89 \\
$s_{true,azimuthal}/s_{true}$ & 0.34 & 0.00 & 0.01 & 0.16 & -0.00 & 0.34 & -0.89 & 1.00 \\
\hline
432 MHz \\
\hline
$r$ & 1.00 & -0.04 & 0.11 & 0.18 & -0.02 & 0.05 & -0.32 & 0.29 \\
$T_{B,true}$ & -0.04 & 1.00 & 0.71 & -0.19 & -0.03 & -0.07 & 0.17 & -0.00 \\
$S_{\nu}$ & 0.11 & 0.71 & 1.00 & 0.35 & 0.04 & -0.03 & 0.03 & -0.00 \\
$s_{true}$ & 0.18 & -0.19 & 0.35 & 1.00 & 0.04 & 0.23 & -0.03 & 0.20 \\
$\theta_{true}$ & -0.02 & -0.03 & 0.04 & 0.04 & 1.00 & -0.01 & -0.01 & -0.02 \\
$(a/b)_{true}$ & 0.05 & -0.07 & -0.03 & 0.23 & -0.01 & 1.00 & -0.18 & 0.35 \\
$s_{true,radial}/s_{true}$ & -0.32 & 0.17 & 0.03 & -0.03 & -0.01 & -0.18 & 1.00 & -0.83 \\
$s_{true,azimuthal}/s_{true}$ & 0.29 & -0.00 & -0.00 & 0.20 & -0.02 & 0.35 & -0.83 & 1.00 \\
\hline
\end{tabular}
\label{tab:rocorr}
%\end{tiny}
\end{minipage}
}
\end{table*}

	Table~\ref{tab:rocorr} displays the Spearman Rank-order cross-correlation coefficients between burst parameters, for bursts observed at 164, 327, and 432 MHz.
	Spearman's Rank-order cross-correlation acts on the {\it ranking} of the values instead of on their actual values, as a regular cross-correlation (e.g. Pearson's) would.
	This has the huge advantage that even a non-linear (polynomial, exponential, power-law, etc.) relationship between parameter x and parameter y can achieve a cross-correlation coefficient near unity.
	The parameters are:
	\begin{itemize}
		%\item $t_{cycle}$ is the time since solar minimum, taken to be about 1995/01/01.
		%\item $t_{noon}$ is the time since local noon. $|t_{noon}|$ is naturally used to search for effects potentially correlated with Earth atmospheric refraction.
		\item $r$ is the angular distance to Sun center.
		\item $T_{B,true}$ is the peak brightness temperature (averaged over 10 s) of the deconvolved gaussian source (as described in Figure~\ref{fig:h_peakintensities} and Section~\ref{sect:size_distribution}).
		\item $S_{\nu}$ is the burst flux (averaged over 10 s).
		\item $s_{true}$ is the FWHM size of the source, deconvolved as explained in Section~\ref{sect:size_distribution}.
		\item $\theta_{true}$ is the orientation of the semi-major axis of the source, from the map x-axis.
		\item $(a/b)_{true}$ is the ratio of the semi-major to semi-minor axis of the source, an easier quantity to visualize than the eccentricity.
		\item $s_{true,radial}/s_{true}$ is a measure of the source's radial elongation, normalized to source rms size.
		\item $s_{true,azimuthal}/s_{true}$ is a measure of the source's azimuthal elongation, normalized to source rms size.
	\end{itemize}

%==========================================================================================================================================================================================================	
%==========================================================================================================================================================================================================	
%==========================================================================================================================================================================================================	
%%\section{Thermal bremsstrahlung and emission energy gradient}\label{appendix:epslocation}
%%
%%	Assuming we are dealing with thermal bremsstrahlung, the following formula applies:
%%
%==================================================================================================================	
\bibliographystyle{apj}
%\bibliography{../psh_biblio}

%+++++++++++++++++++++++++++++++++++++++++++++++++++++++++++++++++++++++++++++++++++++++++++++++++++++++++++++++++++++

%% Included in this acknowledgments section are examples of the
%% AASTeX hypertext markup commands. Use \url without the optional [HREF]
%% argument when you want to print the url directly in the text. Otherwise,
%% use either \url or \anchor, with the HREF as the first argument and the
%% text to be printed in the second.

\acknowledgments

We thank the referee for thoughtful comments on this manuscript.
PSH was supported by NASA Heliospheric Guest Investigator grant NN07AH74G and NASA Contract No. NAS 5-98033.
The NRH is funded by the French Ministry of Education, the French program on Solar Terrestrial Physics (PNST), the Centre National d'Etudes Spatiales (CNES), and the R\'egion Centre.
MV and AK acknowledge support from the CNES and the PNST.
Special thanks to Anne Bouteille, Gordon Hurford, Hugh Hudson, S\"am Krucker, and Jim McTiernan for discussions pertaining to this and other related work.

{\it Facilities:} \facility{Nan\c{c}ay Radioheliograph}.


%% in parentheses, after the keyword, but they will not be verified.

%{\it Facilities:} \facility{RHESSI}, \facility{Hinode (XRT)}, \facility{GOES}, \facility{FOXSI}.



\end{document}